\documentclass[preprint]{aastex}
\usepackage{float,graphicx}
\usepackage{color}
\newcommand{\ri}[1]{\textit{\color{red} #1}}
\newcommand{\bi}[1]{\textit{\color{blue} #1}}

\title{Observational Results of a Multi-Telescope Campaign in Search of Interstellar Urea [(NH$_2$)$_2$CO]}
\author{Anthony J. Remijan}
\affil{National Radio Astronomy Observatory, Charlottesville, VA 22903}
\author{Lewis E. Snyder}
\affil{Department of Astronomy, University of Illinois at Urbana-Champaign, Champaign, IL 61801}
\author{Brett A. McGuire}
\affil{Division of Chemistry and Chemical Engineering, California Institute of Technology, Pasadena, CA 91125}
\author{Hsin-Lun Kuo, Leslie W. Looney, \& Douglas N. Friedel}
\affil{Department of Astronomy, University of Illinois at Urbana-Champaign, Champaign, IL 61801}
\author{G. Yu Golubiatnikov\footnote{Guest worker, 1999-2000: permanent address: Institute of Applied Physics of RAS, Nizhny, Novgorod 603600, Russia} \& Frank J. Lovas}
\affil{Sensor Science Division, National Institute of Standards and Technology, Gaithersburg, MD 20899}
\author{V. V. Ilyushin, E. A. Alekseev, \& S. F. Dyubko}
\affil{Institute of Radio Astronomy of NASU, Chervonopraporna 4, 61002 Kharkov, Ukraine}
\author{Benjamin J. McCall}
\affil{Departments of Chemistry \& Astronomy, University of Illinois at Urbana-Champaign, Champaign, IL 61801}
\author{Jan M. Hollis}
\affil{NASA Goddard Space Flight Center, Greenbelt, MD, 20771}

\begin{document}

\begin{abstract}

In this paper, we present the results of an observational search for gas phase urea [(NH$_2$)$_2$CO] observed towards the Sgr B2(N-LMH) region.  We show data covering urea transitions from $\sim$100 GHz to 250 GHz from five different observational facilities: BIMA, CARMA, the NRAO 12 m telescope, the IRAM 30 m telescope, and SEST.  The results show that the features ascribed to urea can be reproduced across the entire observed bandwidth and all facilities by best fit column density, temperature, and source size parameters which vary by less than a factor of 2 between observations merely by adjusting for telescope-specific parameters.   Interferometric observations show that the emission arising from these transitions is cospatial and compact, consistent with the derived source sizes and emission from a single species. Despite this evidence, the spectral complexity, both of (NH$_2$)$_2$CO and of Sgr B2(N), makes the definitive identification of this molecule challenging. We present observational spectra, laboratory data, and models, and discuss our results in the context of a possible molecular detection of urea.

\end{abstract}

\section{Introduction}
\label{introduction}

Although a large diversity of molecules has been detected in the gas phase in the interstellar medium (ISM), surprisingly few amines or amides, which contain an -NH$_2$ functional group, are among them.  Amines are important for pre-biotic chemistry as amino acids form the building blocks of life on earth (cf. Kim \& Kaiser 2011 and references therein), and simple amines can be made by reactions with ammonia (NH$_3$).  Yet only 5 amines and amides, namely formamide (H$_2$NCHO) \citep{Rubin1971}, cyanamide (H$_2$NCN) \citep{Turner1975}, acetamide (H$_2$NCOCH$_3$) \citep{Hollis2008} , methylamine (H$_3$CNH$_2$) \citep{Fourikis1974}, and amino-acetonitrile (H$_2$NCH$_2$CN) \citep{Belloche2008} have been confirmed in the gas phase toward astronomical environments.  However, recent chemical models and laboratory experiments on the ice analogs of interstellar grain mantles have predicted detectable abundances of potential interstellar amines including hydroxylamine (H$_2$NOH) \citep{Congiu2012,Garrod2008}, methoxyamine (H$_3$CONH$_2$), carbamic acid (H$_2$NCOOH) and urea (H$_2$N)$_2$CO \citep{Garrod2008}. 

Of these molecules, only urea contains a second nitrogen atom - a rarity among known interstellar species.  In fact, of the $\sim$60 N-bearing species, only seven - N$_2$ \citep{Knauth2004}, N$_2$O \citep{Ziurys1994}, N$_2$H$^+$ \citep{Green1974}, H$_2$NCH$_2$CN \citep{Belloche2008}, H$_2$NCN \citep{Turner1975}, HNCNH \citep{McGuire2012}, and $E-$cyanomethanimine (HNCHCN) \citep{Zaleski2013} - contain a second nitrogen.  The characterization and quantification of such doubly-nitrogenated species, like urea, will therefore provide invaluable insight into the mechanisms of nitrogen chemistry in the ISM.  Additionally, due to its importance in prebiotic chemistry, urea's distribution in the ISM may further provide information on the availability of the building blocks of life on young planets \citep{Miller1959}.

The first evidence for extraterrestrial urea was provided in 1975 when it was detected in an analysis of  two samples of the Murchinson C2 chondrite \citep{Hayatsu1975}.  More recently, urea was tentatively detected on grain mantles toward the protostellar source NGC 7538 IRS9 \citep{Raunier2004}.  However, because vibrational transitions in the solid phase are frequently ambiguous, a dedicated search for rotational spectral features of gas phase urea is warranted to both 1) verify its presence in interstellar environments and to 2) better understand the limits of amine formation in the Galaxy.  In this paper, we present observational evidence for gas phase urea observed towards the Sgr B2(N-LMH) region.  We show data covering urea transitions from $\sim$100 GHz to 250 GHz from five different observational facilities.  The spectroscopic parameters of urea are discussed in \S\ref{structure}, the observations are presented in \S\ref{observations}, results are detailed in \S\ref{results}, data analysis procedures are discussed in \S\ref{dataanalysis}, and a discussion follows in \S\ref{discussion}.

\section{Spectroscopic Parameters}
\label{structure}

The first measurements of the microwave spectra of urea were made from 5 GHz to 50 GHz using a heated waveguide cell \citep{Brown1975}. This study provided the rotational analysis, $^{14}$N quadrupole coupling hyperfine structure analysis, and dipole moment determination ($\mu _b=3.83$ D) from Stark effect measurements. Further measurements of the $^{14}$N quadrupole coupling hyperfine structure on transitions below 20 GHz were subsequently reported by Kasten et al. (1986) and Kretschmer et al. (1996). While these literature data provided a firm basis for predicting transitions up to about 100 GHz, the uncertainties were still on the order of 1 MHz for many of the high line strength transitions necessary for interstellar searches. This situation prompted new spectroscopic measurements at NIST over the frequency range from 59 GHz to 114 GHz.  A free space cell was equipped with a heated pulsed nozzle to create a supersonic expansion of neon and urea with rotational temperature about 10 K.  A millimeter-wave synthesizer was employed directly as the radiation source, which passed through the molecular beam and then was focused on a liquid-He-cooled InSb hot-electron bolometer. A total of 38 rotational transitions was measured with Type B, k = 2 (2$\sigma$) uncertainties \citep{Taylor1994} of $<$50 kHz with J ranging from 3 to 10 and K$_a$ from 0 to 5.  No hyperfine structure was resolved in these measurements. Later, the Kharkov group carried out higher frequency measurements to allow for an interstellar search for urea up to the 1 mm region. Using a heated quartz absorption cell utilizing an automated synthesizer-based spectrometer (Ilyushin et al.\ (2005)), the Kharkov group provided 75 new measurements between 78 GHz and 240 GHz.  The urea lines for which we searched were calculated using the millimeter wave data discussed above, as well as the hyperfine-free data from the existing literature cited earlier. 

Table \ref{targettrans} lists the targeted urea transitions in the 1 - 3 mm wavelength range.  These transitions were specifically chosen for an interstellar search because they are the strongest lines in the 1 - 3 mm window with the lowest upper state energies.  We have targeted all transitions in interconnected energy levels so that an identification of urea will be robust. Finally, these particular transitions were selected because given pairs of lines have nearly the same rest frequencies.  For example, the 18$_{1,17}$ - 17$_{2,16}$ and 18$_{2,17}$ - 17$_{1,16}$ transitions have frequencies determined to be 211077.808(31) MHz.  The near frequency coincidence of these transitions effectively doubles the measured intensity of a urea feature at this frequency and increases the chances of an interstellar detection. Throughout the remainder of this paper, the usual shorthand for the quantum numbers of these lines will be employed. For example, instead of using two sets of quantum numbers to identify the urea transition at 254494.539(86) MHz, we will use 23$_{*,23}$ - 22$_{*,22}$ where the $\ast$ will serve as a reminder of the near frequency coincidence of these transitions.  All new spectroscopic data, including full line lists of measured and predicted transitions along with the observed minus calculated frequencies are available in Appendix B, supplemental material and online at www.splatalogue.net \citep{Remijan2008a}.

\section{Observations}
\label{observations}

The Sagittarius B2 molecular cloud (Sgr B2) is located 7.1 kpc away from the Sun and within 300 pc of the Galactic center \citep{Reid1988}. The Sgr B2 complex contains compact hot molecular cores, molecular maser emitting regions, and ultracompact sources of continuum radiation surrounded by larger-scale continuum features, as well as extended molecular material.  The Large Molecular Heimat, a source of compact molecular emission with a spatial extent of $\sim$5\arcsec, resides within the larger Sgr B2(N) molecular envelope  \citep{Hollis2003}. Sgr B2(N) is the preeminent source for the study of large complex interstellar molecules, prompting spectral line surveys towards the region from the centimeter to submillimeter wavelengths (see e.g. Turner et al. 1989; Nummelin et al. 1998; Friedel et al. 2004; Neill et al. 2012).

Observations to identify new astronomical molecules \citep{Remijan2008a,Pulliam2012,McGuire2012,Pety2012,Loomis2013,Zaleski2013} typically rely on single dish measurements; yet interferometric observations provide a significant advantage for identifying low abundance molecules with compact distributions \citep{Belloche2008,Belloche2009}.  The determination of the co-spatial emission across targeted transitions has proven instrumental in the detection and confirmation of low-abundance species \citep{Belloche2008}. As such, acquiring sufficient evidence to support the presence of interstellar urea toward Sgr B2 (N-LMH) required a vast suite of astronomical instrumentation including: the Swedish-ESO Submillimetre Telescope (SEST) (\S3.1),the Berkeley-Illinois-Maryland-Association Array (BIMA)  (\S3.2), the Combined Array for Research in Millimeter-wave Astronomy (CARMA)  (\S3.3), and the NRAO 12 m and IRAM 30 m Telescopes (\S3.4).\footnote{Access to the entire observational dataset for this work is available via the Spectral Line Search Engine accessible at http://www.cv.nrao.edu/$\sim$aremijan/SLiSE/.}   Such a suite of instruments allowed the investigation of the spatial scale of molecular emission from about 1 arcmin to 1 arcsec.  In every observed passband, the continuum level was estimated by choosing channels that are free of line features and, in the case of the array data, then subtracted out in the u-v plane.  Also, unless otherwise highlighed in the specific telescope section, the absolute amplitude uncertainty for each facility is assumed to be accurate to within 20\%.  We briefly describe the scope of each set of observations below.

\subsection{SEST Observations}
\label{sest}

Single dish data with the 15 m telescope were taken from a survey conducted towards Sgr B2(N) (hereafter, the Nummelin Survey) from 218.30 GHz to 263.55 GHz by Nummelin et al. (1998).  The pointing position for these observations was $\alpha$(J2000)= 17h47m19.9s, $\delta$(J2000)= -28\degr 22\arcmin 19.3\arcsec.  The half-power beamwidth (HPBW) ranged from 19\arcsec to 23\arcsec over the survey frequency range. The frequency resolution was 1.4 MHz with resulting velocity resolution of 1.8 km s$^{-1}$ at 230 GHz. The final spectra were resampled to 1.0 MHz channel separation.  The complete observational parameters are described in detail in Nummelin et al.\ (1998).

\subsection{BIMA Array Observations}
\label{bima}

Dedicated searches for transitions of urea toward Sgr B2(N-LMH) were conducted between 2000 November and 2003 October with the BIMA array in C configuration.  The observations were taken toward the Sgr B2 (N-LMH) region with a phase center of (J2000) $\alpha$=17h47m19.92s,  $\delta$=-28\degr22\arcmin18.37\arcsec.  Each spectral window had a bandwidth of 50 MHz and 128 channels, which gave a channel resolution of 0.39 MHz per channel ($\sim$0.5 km s$^{-1}$ at 230 GHz).  The quasar 1733-130 was used to calibrate the antenna based gains.  The absolute amplitude calibration of 1733-130 was based on planetary observations at each observing frequency.  The BIMA array observations covered the urea transitions listed in Table \ref{targettrans} at 91 GHz, 102 GHz, 113 GHz, 211 GHz, 222 GHz, and 232 GHz.

\subsection{CARMA Observations}
\label{carma}

An extended search for urea utilizing CARMA was carried out in June and October 2007 with 14 antennas in C- and D-array configurations and later in June 2008 and November 2009 with 15 antennas in C- and D-array configurations. The phase center was the same as the BIMA observations. All spectral windows had 63 channels with 0.49 MHz spectral resolution ($\sim$0.6 km s$^{-1}$ at 230 GHz).  MWC 349 and 3C454.3 were used as primary flux calibrators and the quasar 1733-130 was used to determine the antenna-based gain solutions. The passbands  were calibrated using the bright astronomical source 3C454.3 for the  wideband (500 MHz) windows and an internal noise source for the spectral (narrowband, $\sim$32 MHz) windows.  The CARMA array observations covered the urea transitions listed in Table 1 at 222 GHz, 232 GHz, 243 GHz, and 254 GHz.

All data from both the BIMA and CARMA array observations were continuum  subtracted, combined, and imaged with the MIRIAD software package \citep{Sault1995}.  To include all the data from a source with multiple tracks or in multiple arrays, the data were inverted in {\it u-v} space.  

\subsection{NRAO 12 m and IRAM 30 m Observations}
\label{nrao}

A 2 mm spectral line survey from 130 GHz to 170 GHz (hereafter, the Turner 2 mm Survey) was conducted with the NRAO 12 m telescope by B. E. Turner between 1993 and 1995 \citep{Remijan2008c}. The HPBW varied from 38\arcsec - 46\arcsec across the band.  The reported pointing position for these observations was $\alpha$(J2000)=17h47m19.29s and $\delta$(J2000)=-28\degr22\arcmin17.3\arcsec.  The hybrid spectrometer was used, providing a bandwidth of 600 MHz across 768 channels for a spectral resolution of 0.781 MHz ($\sim$1.3 km s$^{-1}$ at 150 GHz).  These data were mined for the stronger 2 mm lines of urea listed in Table \ref{targettrans}. Several promising matches to urea emission lines were found, however, the results were inconclusive due to the lower sensitivity of these observations.  We therefore conducted follow-up observations with the IRAM 30 m telescope to achieve the required sensitivity levels. The IRAM 30 m has a HPBW of $\sim$17\arcsec at 150 GHz, which is much smaller than the beam of the 12 m telescope.  The smaller beam greatly reduces beam dilution for compact sources, which increases the measured emission.  The 30 m observations were conducted under marginal weather conditions in September 2009.   The pointing position was $\alpha$(J2000)=17h47m19.92s and $\delta$(J2000)=-28\degr22\arcmin18.37\arcsec.  Two bands of the heterodyne Eight MIxer Receiver (EMIR) connected to the autocorrelation VErsatile SPectrometer Array (VESPA) with two polarizations on each band were used.  The spectral resolution of these observations was 0.320 MHz ($\sim$0.71 km s$^{-1}$ at 135 GHz) over 240 MHz of bandwidth. Several channels were flagged in the center of the band due to birdies caused by the instruments. Pointing was corrected by observing a nearby quasar 1757-240. Data were taken in position-switching mode with a 1\degr offset to the west of the reference position. Data were flagged, calibrated, and analyzed with the GILDAS software (http://www.iram.fr/IRAMFR/GILDAS).   

\section{Results}
\label{results}

\begin{figure}
\centering
\plotone{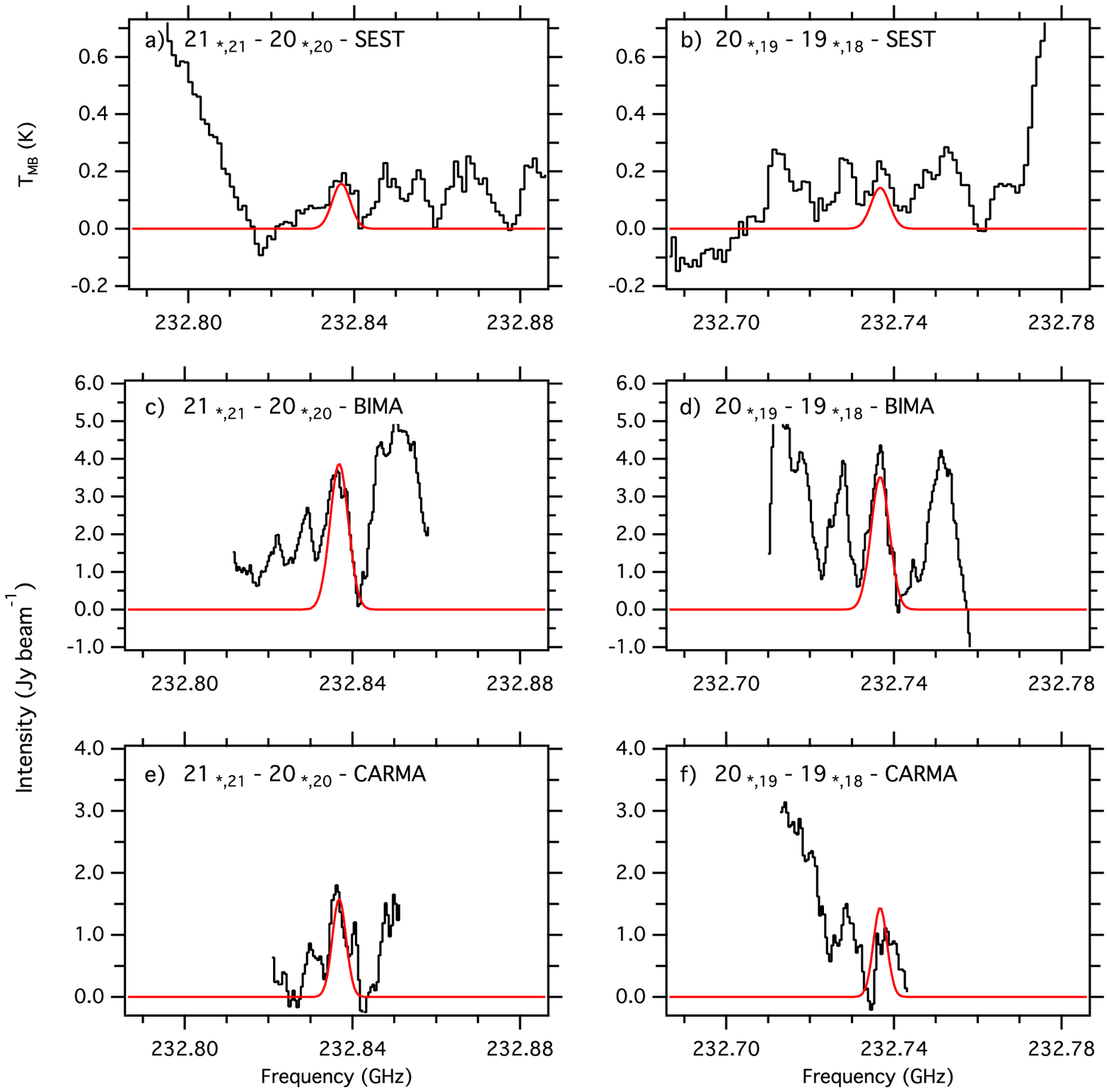}
\caption{$21_{*,21} - 20_{*,20}$ and $20_{*,19} - 19_{*,18}$ transitions in the Nummelin Survey, BIMA observations, and CARMA observations.  The average rms noise measured in the SEST spectra is $\sim$60 mK (Nummelin et al 1998).  The average rms noise in the BIMA and CARMA spectra are $\sim$0.8 and $\sim$0.15 Jy beam$^{-1}$, respectively.  All interferometric spectra are taken at the peak of the emission region and Hanning smoothed over 3 channels for display purposes.  The red Gaussian profile represents the expected intensity and lineshape for the urea transitions based on the best-fit column density and temperature and corrected for telescope and observation-specific parameters (see \S\ref{dataanalysis}).}
\label{2120comp}
\end{figure}

The first indication that urea may be present in the gas phase in Sgr B2(N) came from the report of an unidentified transition in the Nummelin Survey at 232.837 GHz (Figure \ref{2120comp}a).  The reported line is nearly coincident with the 21$_{*,21}$ - 20$_{*,20}$ transition of urea at 232.837 GHz (Table \ref{targettrans}). If the feature at 232.837 GHz is due to urea, then the 20$_{*,19}$ - 19$_{*,18}$ urea transitions at 232.737 GHz must also be present and at a comparable intensity.  

In fact, a feature was identified near 232.737 GHz (Figure \ref{2120comp}b) but was assigned by the authors to the blended J = 14 - 13, K = 0 and 1 transitions of $^{13}$CH$_3$CCH.   However, if the carrier of this feature was in fact $^{13}$CH$_3$CCH, one would expect to see blended lines from the J = 15 - 14, K = 0 and 1 transitions of $^{13}$CH$_3$CCH near 249.359 GHz.  No lines were reported at this frequency and there is no evidence for these transitions by inspecting the data.  Finally, there have been no other transitions of $^{13}$CH$_3$CCH reported toward this source, suggesting that $^{13}$CH$_3$CCH is not the carrier of the 232.737 GHz line reported in the Nummelin Survey. 

A search was conducted to determine the possibility of coincidental overlap with interloper molecules using all available spectroscopic information from the JPL, Lovas/NIST, and CDMS databases.\footnote{Accessible via www.splatalogue.net.}  No reasonable overlapping transitions were found for these or any other features presented here, with the exception of two transitions of methyl formate (CH$_3$OCHO) in the $v$=1 state near the 232 GHz lines.  CH$_3$OCHO has been well-studied and modeled by Belloche et al. (2013) in this source.\footnote{Interferometric observations have also been conducted targeting CH$_3$OCHO in this source by Friedel (2005) which suggest a more compact source size.}  

\begin{figure}
\centering
\plotone{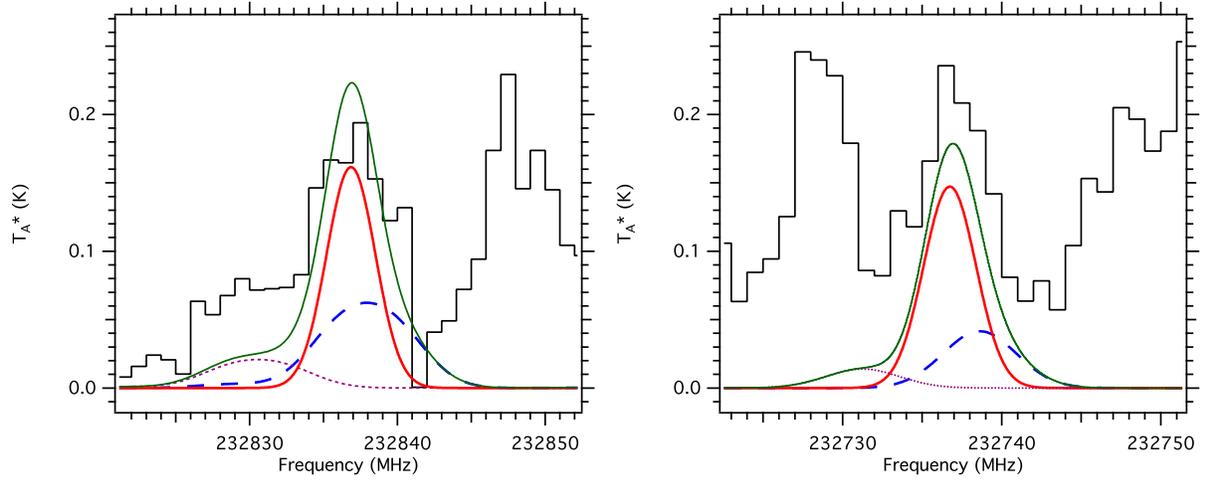}
\caption{$21_{*,21} - 20_{*,20}$ and $20_{*,19} - 19_{*,18}$ transitions in the Nummelin Survey. The best-fit urea model spectra is shown in red over the observations in black.  A simulation of methyl formate emission, using $T_{ex} = 80$ K, $\Delta V = 7$ km s$^{-1}$, and a source size of 4$\arcsec$, as described by Belloche et al. (2013), is shown as a dashed blue line (63.5 km s$^{-1}$ component, $N_T = 4.37 \times 10^{17}$ cm$^{-2}$) and a dotted magenta line (73.5 km s$^{-1}$ component, $N_T = 1.46 \times 10^{17}$ cm$^{-2}$).  A total simulated spectrum including both urea and methyl formate emission is shown as a thin green line.}
\label{fullmodel}
\end{figure}

Based on their derived parameters, we have simulated the contribution to the observed signals from CH$_3$OCHO at local thermodynamic equilibrium (LTE), assuming low optical-depth, following the methods outlined in \S\ref{dataanalysis}.  The simulations are corrected for beam-dilution effects.   The results of the model for the transitions near the 232 GHz urea lines, as well as the relevant parameters used for the simulation, are displayed in Figure \ref{fullmodel}.  The resulting model also well-reproduces unblended CH$_3$OCHO transitions seen in the Nummelin survey.    In both cases, the CH$_3$OCHO emission is insufficient to account for the observed flux.  Additionally, in the case of the 232.737 GHz line, the peak of the CH$_3$OCHO emission is not frequency-coincident with the peak of the observed emission.  Thus, the observed line profile cannot be well reproduced without contribution from a second emission line centered at the frequency of the urea transition.

The interferometric observations with BIMA and CARMA (Figures \ref{2120comp}c-\ref{2120comp}f) show far more intense lines than seen in the single-dish data, compared to adjacent spectral features, indicating that the emission is likely arising from relatively compact material.   If this is the case, the higher spatial resolution of the interferometric observations will provide beam sizes which are more well-matched to the source, reducing the beam dilution which is the likely cause of the weak features observed in the single-dish data.  In fact, some of the emission seen in the BIMA array data is being resolved out by the CARMA observations thus putting better constraints on the size of the emitting region.

\begin{figure}
\centering
\plotone{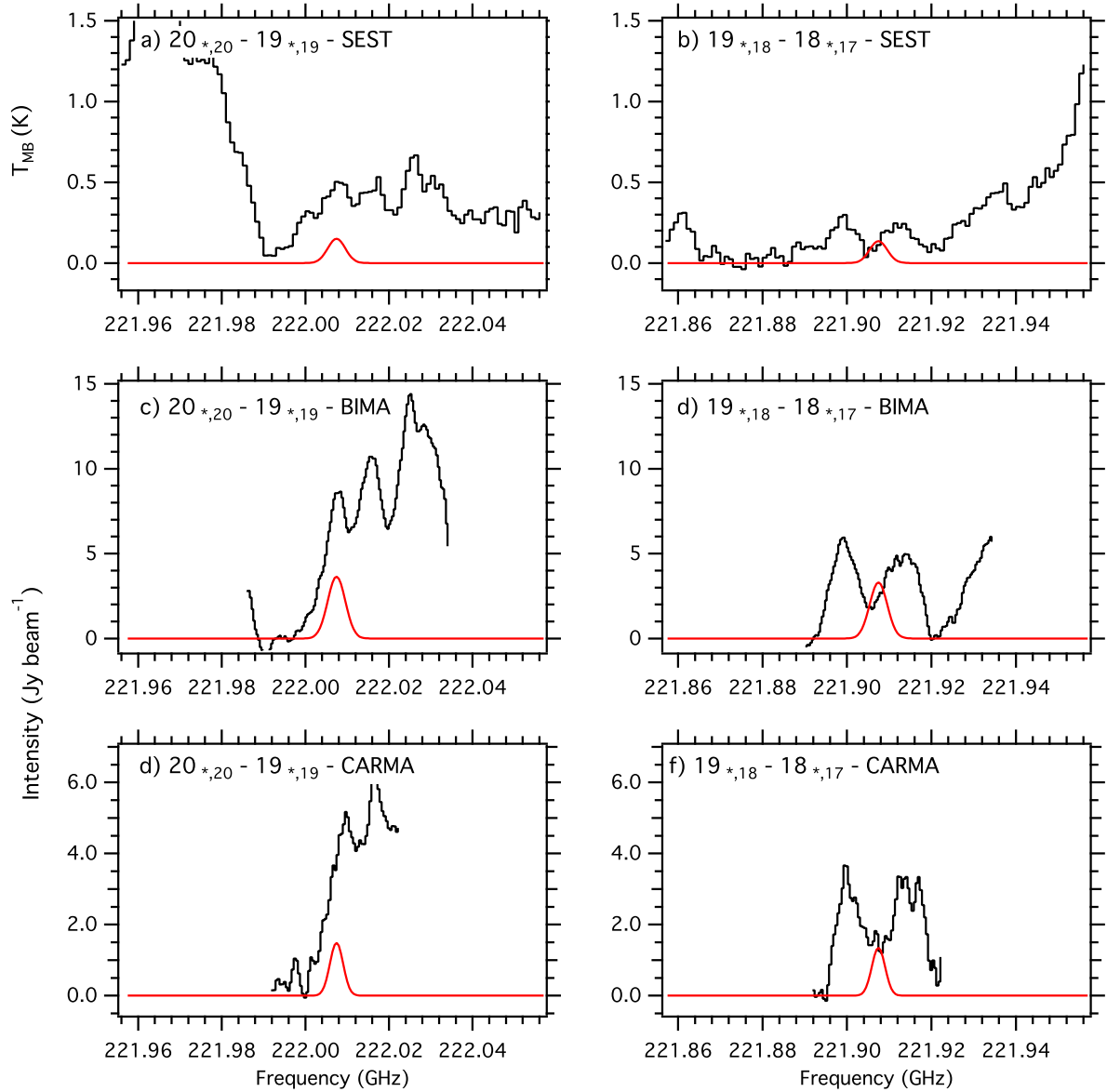}
\caption{$20_{*,20} - 19_{*,19}$ and $19_{*,18} - 18_{*,17}$ transitions in the Nummelin Survey (panels a \& b), BIMA observations (panels c \& d), and CARMA observations (panels e \& f). The rms noise for the SEST data is the same as what is reported in Figure 1.  The average rms noise in the BIMA and CARMA spectra are $\sim$0.7 and $\sim$0.1 Jy beam$^{-1}$, respectively.  The best-fit model spectra is shown in red over the observations in black.}
\label{2221comp}
\end{figure}

\begin{figure}
\centering
\plotone{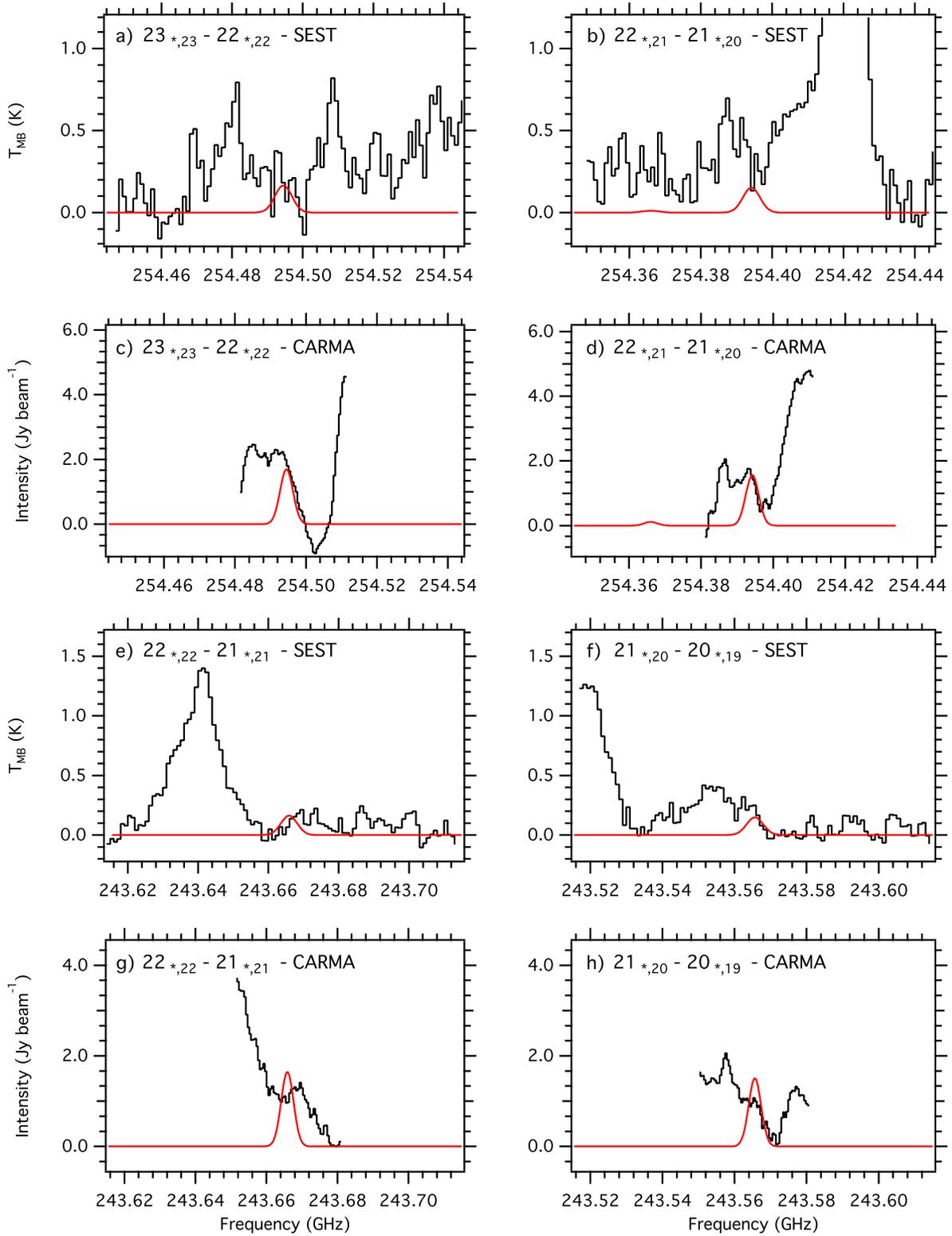}
\caption{Observation of urea transitions at 243 GHz and 254 GHz in the Nummelin Survey and CARMA observations. The rms noise for the SEST data is the same as what is reported in Figure \ref{2120comp} for 243 GHz but closer to $\sim$100 mK at 254GHz.  The average rms noise in the CARMA spectra is $\sim$0.2 Jy beam$^{-1}$.   The best-fit model spectra is shown in red  over the observations in black.}
\label{2322comp}
\end{figure}

Features from the Nummelin Survey are also observed at 222.0 GHz and 221.9 GHz (Figures \ref{2221comp}a and \ref{2221comp}b), corresponding to the $20_{*,20} - 19_{*,19}$ and $19_{*,18} - 18_{*,17}$ transitions of urea, respectively.  While these transitions are likely blended with neighboring transitions, BIMA and CARMA observations provide more convincing evidence, as shown in Figures \ref{2221comp}c-\ref{2221comp}f.  As with the 232.737 GHz and 232.836 GHz features, the intensity of these features is higher in the interferometric data, compared to adjacent spectral features, once again likely indicating compact emission.  Additional evidence is shown in Figure \ref{2322comp}, highlighting urea transitions at 243 GHz and 254 GHz in the Nummelin Survey and CARMA data.  No corresponding BIMA data are available at these frequencies.  Several transitions show evidence in the Gaussian wing profiles, however, the lines are largely blended, though no expected lines are missing from the data.

\begin{figure}
\centering
\plotone{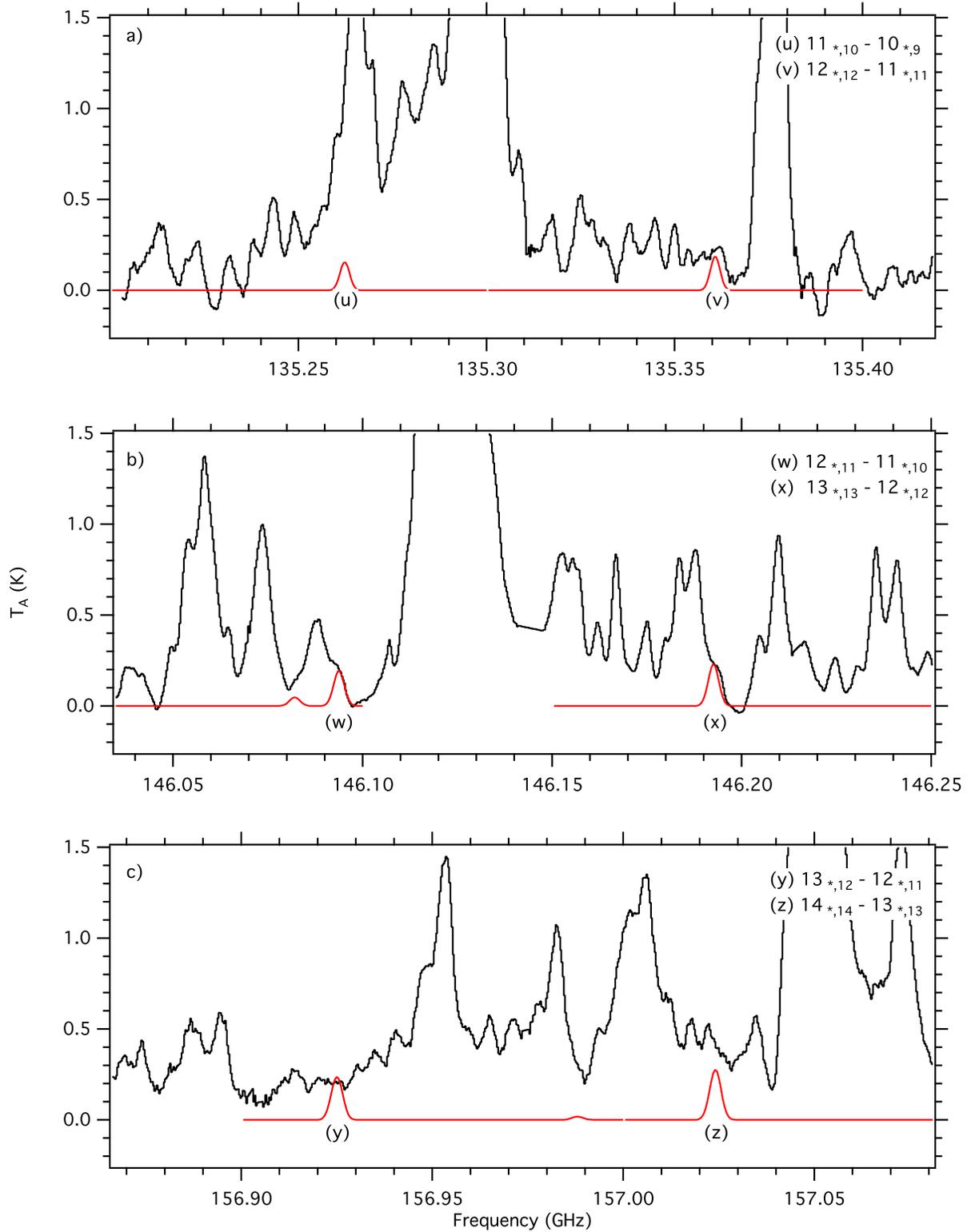}
\caption{Observations of the urea transitions given in Table \ref{targettrans} at 135 GHz, 146 GHz, and 157 GHz using the IRAM 30 m telescope.   The average rms noise measured in the 30 m  spectra is $\sim$16 mK.  The best-fit model spectra is shown in red  over the observations in black.}
\label{30m157}
\end{figure}

\begin{figure}
\centering
\includegraphics[width=6.5in]{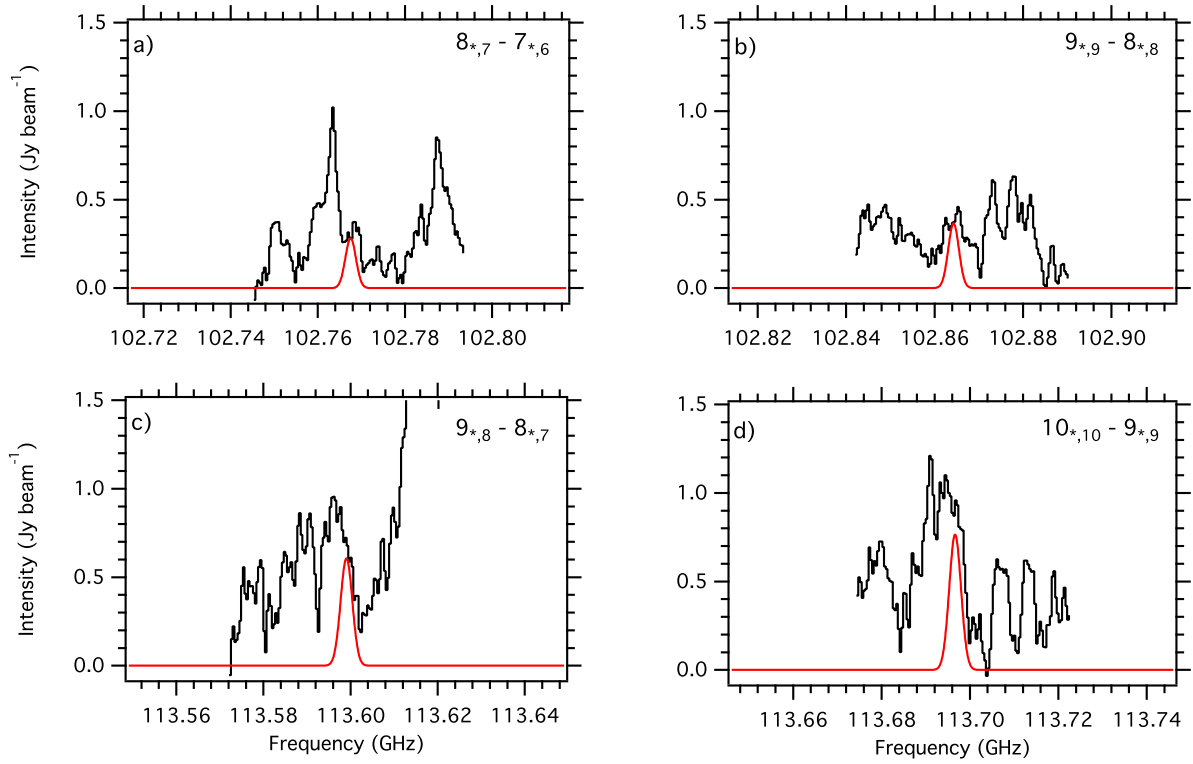}
\caption{BIMA observations of urea at 102 GHz and 113 GHz. The average rms noise in the BIMA spectra are $\sim$0.2 and $\sim$0.3 Jy beam$^{-1}$ for the 102 GHz and 113 GHz data, respectively. The best-fit model spectra is shown in red  over the observations in black. }
\label{bima102}
\end{figure}

Observations of transitions at 135 GHz, 146 GHz, and 157 GHz obtained with the IRAM 30 m telescope, after initial evidence was found in the Turner Survey, are presented in Figure \ref{30m157}.  The $12_{*,12}-11_{*,11}$  transition at 135.36 GHz is moderately well resolved.  As with the features at 221 GHz and 222 GHz, several additional transitions show evidence of well-matched Gaussian wing profiles, and no expected lines are missing from the data.  BIMA observations of the  $7_{*,6} - 6_{*,5}$ and $8_{*,8} - 7_{*,7}$ transitions are obscured by interfering CH$_3^{13}$CN ($J = 5-4$) transitions and H$\alpha$ (41) and He$\alpha$ (41) recombination lines, respectively, and are not presented here.  Figure \ref{bima102} shows the BIMA data for urea transitions at 102 GHz and 113 GHz.  The observed lines are likely highly contaminated with neighboring transitions; however, the observed peak intensities at the urea center frequencies correspond well with those predicted by the model (see \S\ref{dataanalysis}).

\begin{figure}
\centering
\includegraphics[width=6.5in]{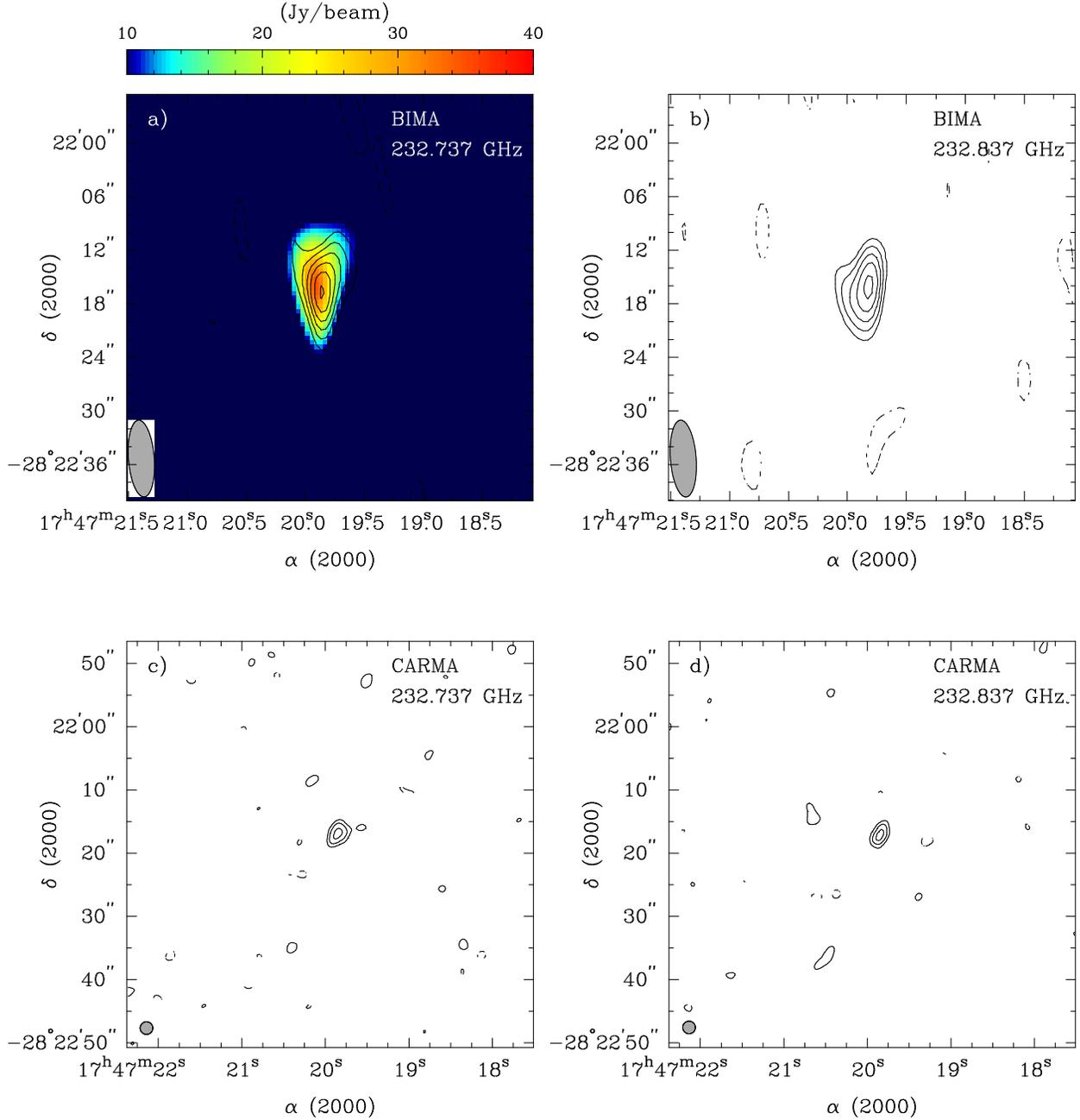}
\caption{(a) (NH$_2$)$_2$CO contours of the $20_{*,19}-19_{*,18}$ transitions overlaid on a Sgr B2N continuum map taken with the BIMA array. The continuum emission maps shown were made from channels which were deemed free from line emission. The contour levels are -3, 3, 4, 5, 6, 7 and 8 $\sigma$ = 0.3 Jy beam$^{-1}$. The continuum unit is Jy beam$^{-1}$. The synthesized beam size is 8.$''$6 x 2.$''$9, shown in the bottom left corner. (b) (NH$_2$)$_2$CO contours of the $21_{*,21}-20_{*,20}$ transition taken with the BIMA array.  Beamsize and countour levels are the same as in (a). (c) (NH$_2$)$_2$CO contours of the $20_{*,19}-19_{*,18}$ transition taken with the CARMA array. The contour levels are -2.5, 2.5, 3.5 and 4.5 $\sigma$ = 0.275 Jy beam$^{-1}$. The synthesized beam size is 1.$''$5 x 1.$''$6, shown in the bottom left corner. (d) (NH$_2$)$_2$CO contours of the $21_{*,21}-20_{*,20}$ transition taken with the CARMA array.  Beamsize and countour levels are the same as in (c).}
\label{bimacarma232}
\end{figure}

\begin{figure}
\centering
\includegraphics{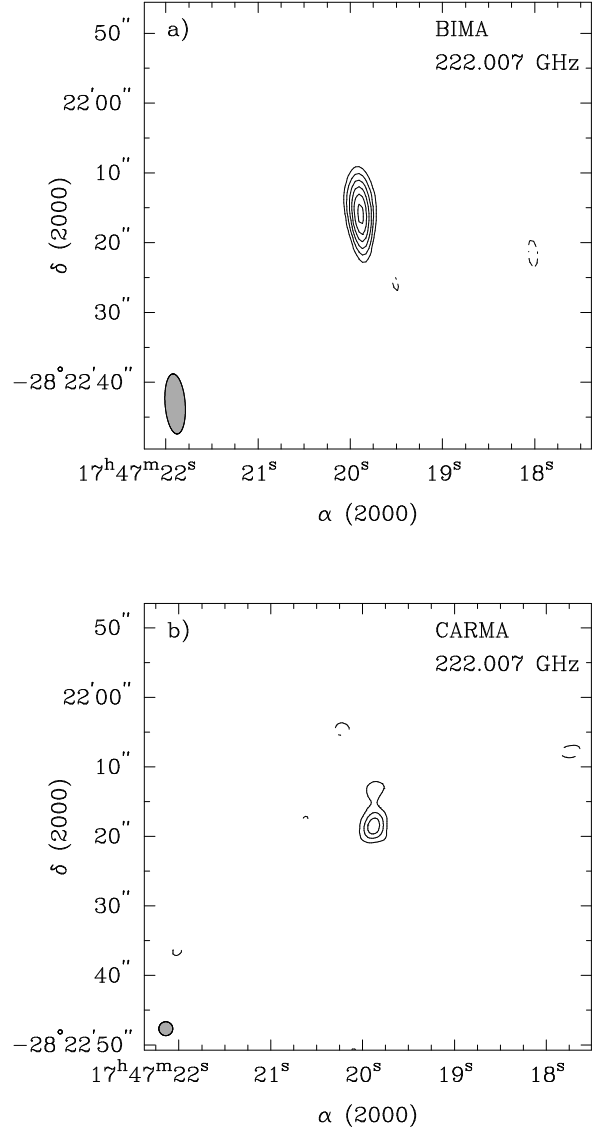}
\caption{(NH$_2$)$_2$CO contours of the $20_{*,20}-19_{*,19}$ transitions taken with the BIMA and CARMA arrays, respectively. (a) The contour levels from the BIMA observations are -3, 3, 4, 5, 6, 7 and 8 $\sigma$ = 0.75 Jy beam$^{-1}$. The synthesized beam size is 8.$''$6 x 2.$''$9, shown in the bottom left corner. (b) The contour levels from the CARMA observations are -2, 2, 4 and 6 $\sigma$ = 3.5 Jy beam$^{-1}$. The synthesized beam size is 1.$''$5 x 1.$''$6, shown in the bottom left corner.}
\label{bimacarma222}
\end{figure}

\begin{figure}
\centering
\plotone{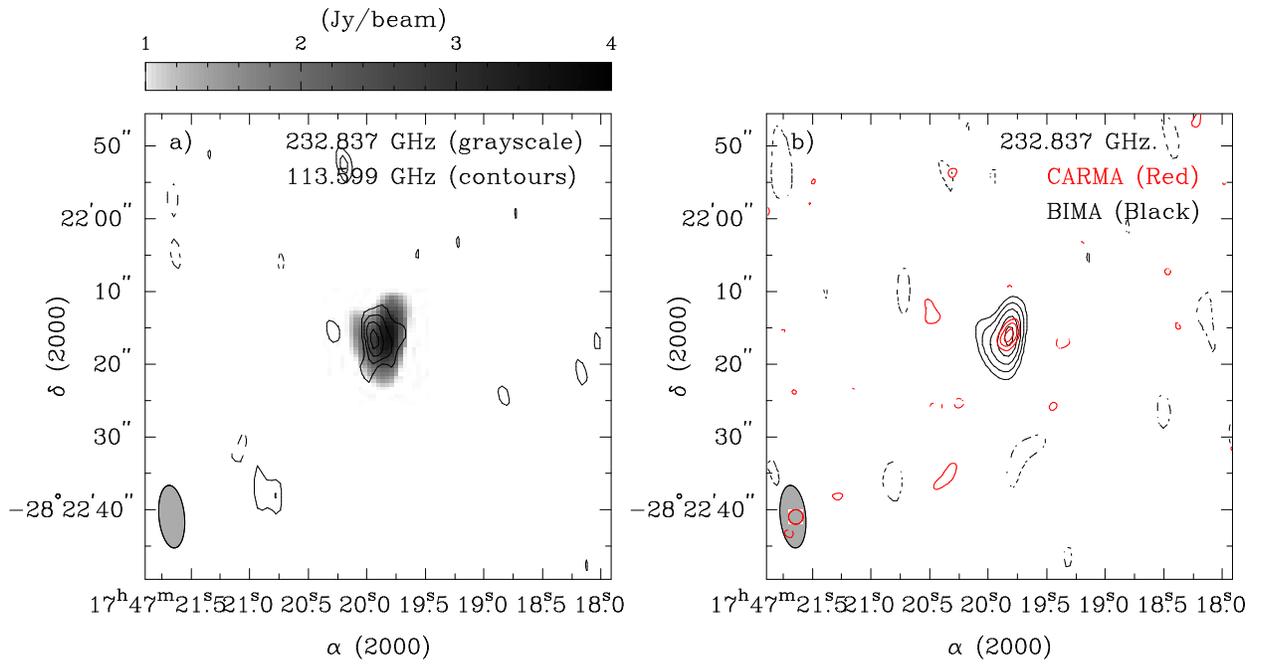}
\caption{(a) $21_{*,21}-20_{*,20}$ (greyscale) and $10_{*,10}-9_{*,9}$ (contour) overlay of BIMA observations.  The contour levels are -3, 3, 4, 5 and 6 $\sigma$ = 0.125 Jy beam$^{-1}$.  1 mm beamsize is the same as in Figure 1a. (b) $21_{*,21}-20_{*,20}$ urea transition observed by CARMA (red contour) and BIMA (black contour).  Beamsize and countour levels are the same as in Figures 1a and 1c, respectively.}
\label{bimacarmacomparison}
\end{figure}

Figure \ref{bimacarma232} shows maps of the 232 GHz transitions of urea with the BIMA and CARMA telescopes.  Panel a) plots the BIMA observation contours over the 232 GHz background continuum.  The urea emission is clearly well-matched to the continuum emission at this frequency.  Both the BIMA and CARMA observations show the urea emission compact and co-spatial at the map center.  This is further highlighted in Figure \ref{bimacarma222}, showing the compact, co-spatial emission of the 222 GHz transition between the two telescopes.  Figure \ref{bimacarmacomparison} shows in panel a) BIMA observations of the 113.6 GHz transition contours overlayed on the 232.8 GHz emission in grayscale again showing the cospatial nature of these high- ($\sim$100 K) and low- ($\sim$30 K) energy states.  Panel b) provides a direct comparison between the CARMA and BIMA observations of the 232.8 GHz transition.  The emission is shown to arise from the same spatial region and is likely to be of the same spatial scale when corrected for the difference in relative beam sizes and resolution. 

Finally, as will be discussed in \S\ref{dataanalysis}, all of the observations can be fit with the same column density, temperature, and source size, conservatively, within a factor of 50\% and a good argument can be made for consistency within the assumed absolute calibration uncertainty of $\sim$20\%.  Additionally, a visual inspection of the modeled profiles for urea transitions computed from the best fits, and compared with the observations, shows that all of the features assigned to urea are reproduced within a factor of $<$50\% in intensity and linewidth.  This agreement, across a broad range in bandwidth, several observational facilities, and both single-dish and interferometric observations, is remarkable, and suggests the possible presence of urea in this source.  Further supporting evidence for all transitions arising from a single carrier is provided by the Student t-test given in Appendix A.\footnote{All spectra presented here are accessible at http://www.cv.nrao.edu/$\sim$aremijan/SLiSE/.  All spectroscopic data used in the assignment are accessible via www.splatalogue.net.}$^,$\footnote{A comprehensive assignment of the additional transitions presented in each urea passband was not completed.}  \textbf{Despite this body of evidence, the spectral complexity of both urea and of Sgr B2(N) make a definitive identification challenging, and require a rigorous, methodical treatment of the data.} 

\section{Data Analysis}
\label{dataanalysis}

Given the varying telescopes used in these observations, it was essential to arrive at consistent determinations of column densities and temperatures.  Meaning, if the assumptions of the physical conditions that give rise to emission features of urea are correct, all that should need to change is the telescope specific parameters in the determination of the relative intensities of the urea features. Also, trying to make an independent Gaussian fit to the integrated intensity of the detected features to compare to the predicted integrated intensity from the model will provide little information for comparison.  This is because several of the features are blended with unknown transitions of other molecules given the high line density of the SgrB2N region.  Interested readers are encouraged to download the spectroscopic data to explore these transitions for detailed spectroscopic modeling and line fitting.$^4$

In order to determine the relative intensity of each urea transition observed between the telescopes, we assume uniform physical conditions, that the populations of the energy levels can be characterized by a Boltzmann distribution, and finally, that the emission is optically thin. Assuming that the molecular species is in LTE and low optical depth, the total beam-averaged column density for an emission line detected by a single dish telescope is given by Equation \ref{sdltecd} \citep{Remijan2005}. 

\begin{equation}
\label{sdltecd}
N_T=\frac{\frac{1}{2}Qe^{E_u/T_r}\Delta T_A^* \Delta V \sqrt{\frac{\pi}{ln 2}}}{\frac{8\pi ^3}{3k}B\nu S \mu ^2 \eta _B (1 - \frac{e^{h\nu/kT_r -1}}{e^{h\nu/kT_{bg} -1}})}
\end{equation}

The line shapes are assumed to be Gaussian,  $\eta _B$ is the telescope beam efficiency, $T_r$ is the rotational temperature,  $\Delta T_A^* \Delta V$ is the product of the fitted line intensity (mK) and line width (km s$^{-1}$), $Q$ is the rotational partition function\footnote{Calculated using the approximation given by Gordy \& Cook (1984), Eq. 3.69} given as 6.7$\cdot T_r^{1.5}$, $S \mu ^2$ is the product of the transition line strength and the square of the dipole moment (Debye$^2$), $E_u$ is the upper rotational energy level (K), B is the beam filling factor given in Equation \ref{beamfilling} where $\Theta_b$ is the circular Gaussian telescope beam size and $\Theta_s$ is the circular Gaussian source size (see Eq 28 of Ulich \& Haas 1976), $\nu$ is the transition frequency (MHz), and $T_{bg}$  = 2.7 K is the cosmic background temperature.

\begin{equation}
\label{beamfilling}
B=\frac{\Theta_s^2}{\Theta_b^2 + \Theta_s^2}
\end{equation}

For interferometric observations, the total beam-averaged column density is Equation \ref{intltecd} where $\Omega _b$ is the solid angle of the beam (square arcseconds), $\int I_v dv$ is the integral of the line intensity (Jy/beam) over velocity (km s$^{-1}$), $\nu$ is given in units of GHz, and the remaining variables are as in Equation \ref{sdltecd} \citep{Miao1995}.  
\begin{equation}
\label{intltecd}
N_T=2.04\times 10^{20} \times \frac{\int  I_v dv Qe^{E_u/T_r}}{B\Omega _b \nu ^3 S \mu ^2}
\end{equation}

Excluding telescope-specific parameters (e.g. interferometer resolving out extended, smooth structures) and molecule-specific parameters, the intensity of observed transitions is determined by the column density, temperature, and molecular emission source size.  Remijan (2003) derived an initial column density and temperature of urea of 3.4$\pm2.0\times 10^{15}$ cm$^{-2}$ and 77$\pm23$ K based on a rotation diagram analysis (excluding the absolute amplitude uncertainty) using the BIMA data at 102, 113, 211, 222, and 232 GHz, however, computational errors in the values of $Q_r$ and $S\mu ^2$ led to an overestimate of the column density.    

Here, a zeroth-order fit of the data found a column density and temperature of $\sim$$8 \times 10^{14}$ cm$^{-2}$ and $\sim$80 K, with source sizes between 2$\arcsec$ and 3.25$\arcsec$ best reproduced the observational results in all cases except for the BIMA 112 GHz transition, which is best fit by a column density of  $\sim1 \times 10^{15}$ cm$^{-2}$.  A predicted model spectrum of urea was then generated based on each of these values, accounting for telescope-specific parameters in each case.  The results of this prediction are detailed in Table \ref{models}.  The results of this fit show that all of the observations, across all of the facilities, can be fit with the same column density, temperature, and source size within a factor of $<$50\%.

Furthermore, the model can provide predictions on peak line intensities for a given beam size for urea transitions which fall within the observable range of several facilities, but which were not observed for this work.  The purpose of these predictions is to guide analysis of archive data or future efforts in the observational identification of additional urea transitions.  These include predictions for the newly operational ALMA facility, which is ideally suited for high-sensitivity observations of compact objects.

Single dish spectra observed with a large beam usually lack spatial information. There is no way to pinpoint where the flux originates from within the beam. Normally this is not a problem for identifying molecules with resolved identifiable spectral lines from a typical source size (i.e. $> 5\arcsec$). However, as suggested by previous SEST and BIMA data, urea is a compact source with weak blended lines. By mapping the transitions with a high-resolution interferometer, we could determine whether they originate from the same region or molecular origin. Good spatial correlations between the channel maps is not a sufficient but rather a necessary condition for identifying urea since bad correlations would suggest molecules of different distributions as sources of these line emissions. How do we quantify the relationships of spatial distribution?  Adapting the quantitative method of Turner \& Thaddeus (1977), a correlation analysis is preferred because we are mainly interested in knowing the association between the lines.  Such an analysis was performed and is detailed in Appendix A.

\section{Discussion}
\label{discussion}

The number of known interstellar molecules is growing at a steady pace, and the number of simple molecules that likely remain undetected is shrinking.  As the field turns to examining more and more complex molecules, the spectral complexity of these molecules is an increasingly frustrating problem.  While such complex spectra provide for the possibility of unambiguous detection over a broad range in frequency space, the corresponding decrease in spectral intensity drastically increases the sensitivity required for detection.  Further, because such molecules tend to co-exist in favorable environments, the resulting observations typically suffer from high line-density and line-confusion.  Consequently, new detections of complex molecules going forward will necessarily have to deal with most or all molecular lines being blended to a lesser or greater degree with other molecules (e.g. Tercero et al. 2013).  The additional confusion caused by multiple velocity components within prototypical sources, such as Sgr B2(N), compounds these issues further.

Here, it has been demonstrated that substantial evidence for an interstellar detection can still be compiled despite a preponderance of blended lines due to overlap with other species and multiple velocity components.  Historically, single-dish observations have been the primary source of new interstellar detections.  This is due largely to the ever increasing spectral sensitivity, high-resolution, and wide bandwidth achievable by such facilities, combined with their sensitivity to both compact and extended structure in sources.  In this work, the broadband and high-sensitivity nature of the single-dish observations provided a wealth of initial signals at frequencies suggesting urea as a carrier.  Moving forward, however, the use of single-dish observations as the sole means for the detection of complex molecules such as urea is unsustainable.  

Interferometric observations provide a number of benefits, but typically trade spectral bandwidth for resolution, making blind, wideband searches for molecules unrealistic.  This is changing with the advent of facilities employing broadband correlators such as the VLA, CARMA, PDBI, and ALMA, but the initial evidence must still likely be conducted with single-dish observatories.  Once a potential carrier is identified via single-dish observations, interferometric observations are the logical next step.  The use of telescope arrays allows for greater spatial resolution, providing better beam coupling to compact sources as well as spatial filtering of velocity components.  By spatially resolving individual, or smaller groups of velocity components, the resulting line confusing in observed spectra is immediately reduced, thus simplifying identifications.  Further, by better coupling to compact sources, signals arising from these areas are less affected by the beam dilution of single-dish measurements, increasing the achievable spectral sensitivity.  Finally, by mapping the spatial extent of a number of potential transitions, we can determine whether the signals arise from the same physical environment, further increasing the likelihood that they arise from the same molecule \citep{Neill2011}.

The criteria for a firm interstellar molecular detection have been well established by Snyder et al. (2005) in response to the reported detection of interstellar glycine by Kuan et al. (2003).  Here, we discuss each point outlined by Snyder et al. (2005) and apply them to our observations of urea.

\textit{Accurate laboratory rest frequencies are required for comparison to observed transitions}.  Quantum mechanical predictions of rotational transitions, even for fairly rigid species, very often lack sufficient accuracy for comparison to astronomical spectra.  In the recent case of glycolaldehyde, a relatively rigid molecule, predictions based on high-accuracy laboratory measurements were found to be in error by factors as large as 15 MHz upon further laboratory measurements \citep{Carroll2010}.  For more spectrally complex molecules, these errors will likely be far greater in magnitude. In this study, all assigned transitions of urea have been measured to high accuracy in the laboratory ($<0.25$ channel widths at 102 GHz).

\textit{Observed spectral features should display constant velocities consistent with known source LSR velocities and should arise from consistent spatial regions.}  A careful analysis of our data indicates that the velocity of the urea transitions is 65.2 km s$^{-1}$ rather than the systematic velocity of 64 km s$^{-1}$ observed in many other molecules.  This velocity is consistent, however, across the observed features, and is a minor deviation from the systematic velocity, which satisfies the first half of this requirement.  Furthermore, inspection of the CARMA and BIMA maps of observed emission at urea frequencies clearly shows such emission arising from the same spatial region, satisfying the second half of the requirement. 

\textit{Model spectra used for comparison must account for all telescope-specific parameters.}  Perhaps the most important of these parameters is beam size and resulting beam-dilution factors.  Because the observations include interferometric measurements, an accurate determination of the source size, coupled with known telescope parameters, allowed for accurate determination of the beam dilution factors involved.  As outlined in \S\ref{dataanalysis}, a model was generated which correctly accounts for all telescope-specific factors, including beam dilution, for use in the analysis.   When compared to observations, this single model reproduces the observations within an intensity factor of $<$50\% across all of our observations.  

\textit{All transitions which are reasonably predicted to have observable intensity must be present or otherwise accounted for}.   Although the majority of the observed transitions are blended to some degree with neighboring features, all transitions that are predicted to be observable by our model are present within a factor of $<$50\% in intensity.  

The statistical test (detailed in Appendix A), focusing on line maps rather than spectra, provides an unconventional approach when most of the lines of interest are $\leq3\sigma$.  This approach is useful for compact sources that could be mapped with an interferometer.  As demonstrated in Friedel \& Snyder (2008), molecules with different formation routes can have very different distributions.  If the weak lines are frequency coincidences resulting from other molecular lines, it is not likely they would all show similar distribution.  Because no source model was assumed here and only simple correlation coefficients are calculated, this can be applied fairly easily to prevent false identifications.  In fact, this technique can easily be applied to facilitate more efficient data mining for connected energy level transitions in large datasets such as those beginning to be produced from ALMA observations. 

On the other hand, no matter how high the confidence intervals are, this analysis does not guarantee signal arises from the same molecular origin.  There are species that are well correlated with each other physically and chemically in the hot core region due to similar formation routes.  As mentioned earlier, there could be a correlation up to 99\% between molecules of compact and extended distribution given their contours peak around the same region.  Therefore, this methodology must be used cautiously, as it can only prove a negative, rather than confirming a positive.

\section{Conclusions}
\label{conclusion}

We have conducted and compiled observations across five different facilities over 150 GHz in bandwidth in an attempt to detect interstellar urea.  Here, we present the resulting evidence for interstellar urea in the gas phase from the observations.  The results show that the features ascribed to urea can be reproduced across the entire observed bandwidth and all facilities by best fit column density, temperature, and source size parameters which vary by less than a factor of 2 between observations merely by adjusting for telescope-specific parameters.  Further, the predicted profiles of the urea transitions resulting from these best fit values reproduce the observed spectra with a factor of  $<$50\% across the entire dataset.  Interferometric observations show that the emission arising from these transitions is cospatial and compact, consistent with the derived source sizes and emission from a single species.  We have discussed and satisfied the essential conditions for the detection of a new molecule as prescribed by Snyder et al.\ (2005).  Finally, we extend our model predictions to include expected linewidths and peak intensities for urea transitions which fall within the operational range of millimeter observatories as a guide to the detection and assignment of additional urea transitions.

Taken as a whole, the results of this observational campaign present tantalizing, but not definitive evidence for the presence of urea in this source.  This work highlights both the difficulties of identifying new interstellar molecules which have complex, low-intensity signals in a source with high-line density spectra and the great care which must be taken to treat such detections correctly.  Indeed, single-dish observations of such molecules alone are likely to be increasingly insufficient moving forward.  The power of interferometric observations to provide evidence of co-spatial emission, as well as to decrease spectral confusion, will be essential to future detections of complex species.

The methodology used here, if perhaps not to this extent, is not new. Two recent examples, those of aminoacetonitrile and ethyl formate, rely on essentially the same analysis methodology with similar data and arguments presented here.  Yet, even with the body of evidence we have presented, we cannot claim a definitive detection of urea because of the extreme spectral complexity of both the source and the molecule.  Future detections in these crowded spectral regions will likely become only more difficult, and great care will need to be taken to claim a definitive detection, especially of large complex organic molecules in the mm and sub-mm regions.  Instead, observations of these sources at lower frequencies, where spectral density is much lower, have recently proven quite fruitful in this regard (see, e.g. Neill et al. 2012, McGuire et al. 2012, Loomis et al. 2013, Zaleski et al. 2013) and may remain one of the only ways to unambiguously identify large molecular species in these line-rich sources.

\acknowledgments

We thank the anonymous referee for very helpful comments which improved the quality of this manuscript.  B.A.M. gratefully acknowledges G. A. Blake for his support, and funding by an NSF Graduate Research Fellowship.  H.-L.K. and B.J.M. gratefully acknowledge funding by a UIUC Critical Research Initiative. We acknowledge support from the Laboratory for Astronomical Imaging at the University of Illinois and NSF AST 99-81363 and AST 02-28953. Support for CARMA construction was derived from the states of California, Illinois, and Maryland, the James S. McDonnell Foundation, the Gordon and Betty Moore Foundation, the Kenneth T. and Eileen L. Norris Foundation, the University of Chicago, the Associates of the California Institute of Technology, and the National Science Foundation. Ongoing CARMA development and operations are supported by the National Science Foundation under a cooperative agreement, and by the CARMA partner universities. The National Radio Astronomy Observatory is a facility of the National Science Foundation operated under cooperative agreement by Associated Universities, Inc.

\clearpage

\clearpage

\begin{deluxetable}{lccc}
\tablecolumns{4}
\tablecaption{Quantum numbers, calculated frequency, line strength, and upper state energy for 52 targeted transitions of urea}
\tablewidth{0pt}
\tablehead{
\colhead{Transition} & \colhead{Calculated Frequency\tablenotemark{a}} & \colhead{$S_{i,j}$} & \colhead{E$_{u}$}\\
\colhead{} & \colhead{(MHz)} & \colhead{} & \colhead{(K)}
}
\startdata
7$_{1,6}$-6$_{2,5}$ & 91 936.245(10) & 5.430 & 12.22\\ 
7$_{2,6}$-6$_{1,5}$ & 91 936.568(10) & 5.430 & 12.22\\ 
8$_{0,8}$-7$_{1,7}$ & 92 031.812(11) & 7.455 & 15.14\\
8$_{1,8}$-7$_{0,7}$ & 92 031.812(11) & 7.455 & 15.14\smallskip\\

8$_{1,7}$-7$_{2,6}$  & 102 767.524(10) & 6.4242 & 14.17\\
8$_{2,7}$-7$_{1,6}$  & 102 767.544(10) & 6.4243 & 14.17\\
9$_{0,9}$-8$_{1,8}$  & 102 864.308(12) & 8.4553 & 17.10\\
9$_{1,9}$-8$_{0,8}$  & 102 864.308(12) & 8.4553 & 17.10\smallskip\\

9$_{1,8}$-8$_{2,7}$  & 113 599.065(11) & 7.4192 & 16.12\\
9$_{2,8}$-8$_{1,7}$  & 113 599.066(11) & 7.4192 & 16.12\\
10$_{0,10}$-9$_{1,9}$  & 113 696.657(12) & 9.4556 & 19.06\\
10$_{1,10}$-9$_{0,9}$  & 113 696.657(12) & 9.4556 & 19.06\smallskip\\

11$_{1,10}$-10$_{2, 9}$ & 135 262.288(12) &  9.4116 & 20.03\\ 
11$_{2,10}$-10$_{1, 9}$ & 135 262.288(12) &  9.4116 & 20.03\\ 
12$_{0,12}$-11$_{1,11}$ & 135 360.816(14) &  11.456 & 22.98\\ 
12$_{1,12}$-11$_{0,11}$ & 135 360.816(14) &  11.456 & 22.98\smallskip\\ 

12$_{1,11}$-11$_{2,10}$ & 146 093.756(13) & 10.4086 & 21.98\\ 
12$_{2,11}$-11$_{1,10}$ & 146 093.756(13) & 10.4086 & 21.98\\ 
13$_{1,13}$-12$_{0,12}$ & 146 192.586(15) & 12.4562 & 24.93\\ 
13$_{0,13}$-12$_{1,12}$ & 146 192.586(15) & 12.4562 & 24.93\smallskip\\ 

13$_{1,12}$-12$_{2,11}$ &  156 925.049(13) & 11.4060 & 23.94 \\ 
13$_{2,12}$-12$_{1,11}$ &  156 925.049(13) & 11.4060 & 23.94\\ 
14$_{0,14}$-13$_{1,13}$ &  157 024.124(16) & 13.4563 & 26.89\\ 
14$_{1,14}$-13$_{0,13}$ &  157 024.124(16) & 13.4563 & 26.89\smallskip\\ 

18$_{1,17}$-17$_{2,16}$ &  211 077.802(27) & 16.3971 & 33.72\\
18$_{2,17}$-17$_{1,16}$ &  211 077.802(27) & 16.3971 & 33.72\\
19$_{1,19}$-18$_{0,18}$ &  211 177.706(36) & 18.4566 & 36.68\\ 
19$_{0,19}$-18$_{1,18}$ &  211 177.706(36) & 18.4566 & 36.68\smallskip\\ 


19$_{1,18}$-18$_{2,17}$  & 221 907.446(32) & 17.3958 & 35.67\\
19$_{2,18}$-18$_{1,17}$  & 221 907.446(32) & 17.3958 & 35.67\\
20$_{0,20}$-19$_{1,19}$  & 222 007.476(43) & 19.4566 & 38.64\\
20$_{1,20}$-19$_{0,19}$  & 222 007.476(43) & 19.4566 & 38.64\smallskip\\


20$_{1,19}$-19$_{2,18}$  & 232 736.740(39) & 18.3946 & 37.63\\
20$_{2,19}$-19$_{1,18}$  & 232 736.740(39) & 18.3946 & 37.63\\
21$_{0,21}$-20$_{1,20}$  & 232 836.891(51) & 20.4567 & 40.60\\
21$_{1,21}$-20$_{0,20}$  & 232 836.891(51) & 20.4567 & 40.60\smallskip\\

21$_{1,20}$-20$_{2,19}$  & 243 565.666(46) & 19.3936 & 39.59\\
21$_{2,20}$-20$_{1,19}$  & 243 565.666(46) & 19.3936 & 39.59\\
22$_{0,22}$-21$_{1,21}$  & 243 665.931(60) & 21.4567 & 42.56\\
22$_{1,22}$-21$_{0,21}$  & 243 665.931(60) & 21.4567 & 42.56\smallskip\\

22$_{1,21}$-21$_{2,20}$  & 254 394.205(55) & 20.3926 & 41.54\\
22$_{2,21}$-21$_{1,20}$  & 254 394.205(55) & 20.3926 & 41.54\\
23$_{0,23}$-22$_{1,22}$  & 254 494.580(70) & 22.4567 & 44.52\\
23$_{1,23}$-22$_{0,22}$  & 254 494.580(70) & 22.4567 & 44.52\smallskip\\
\enddata
\tablenotetext{a}{2$\sigma$ frequency uncertainties are given in parentheses in units of the last digit and are of type A (coverage factor k = 2) (Taylor \& Kuyatt 1994).} 
\label{targettrans}
\end{deluxetable}

\begin{deluxetable}{ccccccccccccccc}
\tablecolumns{15}
\rotate
\tablecaption{Best fit model parameters for observed transitions of urea.}
\tablewidth{0pt}
\tablehead{
							&							&					\multicolumn{3}{c}{BIMA\tablenotemark{a,d}}							&	&						\multicolumn{3}{c}{CARMA\tablenotemark{a}}											&	&		\multicolumn{2}{c}{SEST\tablenotemark{b}}				&	&		\multicolumn{2}{c}{IRAM\tablenotemark{c}}		           		\\	
\cline{3-5}\cline{7-9}\cline{11-12}\cline{14-15}
							&		\colhead{Frequency}	&		\colhead{$\theta_s$}		&	\colhead{$\theta_b$}	&	\colhead{I}				&	&	\colhead{$\theta_s$\tablenotemark{e}}		&	\colhead{$\theta_b$\tablenotemark{e}}	&	\colhead{I}			&	&	\colhead{$\theta_b$}		&	\colhead{I}				& 	&	\colhead{$\theta_b$}					&	\colhead{I} 		 \\
	\colhead{Transition}			&	\colhead{(MHz)}			&		($\arcsec$)			&	($\arcsec$)			&  		(Jy bm$^{-1}$)			&	&	($\arcsec$)							&	($\arcsec$)						&  	(Jy bm$^{-1}$)			&	&	($\arcsec$)				&	(K)						&	&	($\arcsec$)							&	(K)\tablenotemark{f}			
}
\startdata
$8_{*,7}-7_{*,6}$ 				& 102 767.523 					&  		19					& 	6 					&	0.31		 				&  	& 		\ri{4}								& 	\ri{2} 								& 	\ri{0.11}				& 	&	\nodata					&	\nodata					& 	&  \ri{20.2}									& 	\ri{0.06}					 \\
$9_{*,9}-8_{*,8}$		& 102 864.309 &		19	&	6	&	0.40		 &  &		\ri{4}		&	\ri{2}		&	\ri{0.14}		& &	\nodata	&	\nodata	&  &	\ri{20.2}	&	\ri{0.08}		\smallskip		 \\

$9_{*,8}-8_{*,7}$ 		& 113 599.061 &		8	&	3	&	0.44	 	&  &		\ri{3.7}	&	\ri{2}		&	\ri{0.15}		& &	\nodata	&	\nodata	&  &	\ri{18.1}	&	\ri{0.09}					 \\
$10_{*,10}-9_{*,9}$ 		& 113 696.656 &		8	&	3	&	0.56	 	&  &		\ri{3.7}	&	\ri{2}		&	\ri{0.20}		& &	\nodata	&	\nodata	&  &	\ri{18.1}	&	\ri{0.11}		\smallskip		 \\

$11_{*,10}-10_{*,9}$	& 135 262.296 &		\bi{2}	&	\bi{2}	&	\bi{0.22}	 &  &	\nodata		&	\nodata	&	\nodata		& &	\nodata	&	\nodata	&  &	15.2		&	0.16						 \\
$12_{*,12}-11_{*,11}$	& 135 360.814 &		\bi{2}	&	\bi{2}	&	\bi{0.26}	 &  &	\nodata		&	\nodata	&	\nodata		& &	\nodata	&	\nodata	&  &	15.2		&	0.20			\smallskip		 \\
	
$12_{*,11}-11_{*,10}$	& 146 093.764 &		\bi{2}	&	\bi{2}	&	\bi{0.28}	 &  &	\nodata		&	\nodata	&	\nodata		& &	\nodata	&	\nodata	&  &	14.1		&	0.20			 			 \\
$13_{*,13}-12_{*,12}$	& 146 192.582 &		\bi{2}	&	\bi{2}	&	\bi{0.34}	 &  &	\nodata		&	\nodata	&	\nodata		& &	\nodata	&	\nodata	&  &	14.1		&	0.24			\smallskip		 \\
	
$13_{*,12}-12_{*,11}$	& 156 925.058 &		\bi{2}	&	\bi{2}	&	\bi{0.35}	 &  &	\nodata		&	\nodata	&	\nodata		& &	\nodata	&	\nodata	&  &	13.1		&	0.26						 \\
$14_{*,14}-13_{*,13}$	& 157 024.119 &		\bi{2}	&	\bi{2}	&	\bi{0.40}	 &  &	\nodata		&	\nodata	&	\nodata		& &	\nodata	&	\nodata	&  &	13.1		&	0.29			\smallskip		  \\	

$19_{*,18}-18_{*,17}$ 	& 221 907.393 &		9	&	3	&	3.16		 &  &		2.7		&	2.1		&	1.05			& &	18.6		&	0.12		&  &	\ri{9.2}	&	\ri{0.52}					 \\	
$20_{*,20}-19_{*,19}$	& 222 007.468 &		9	&	3	&	3.48		 &  &		2.7		&	2.1		&	1.16			& &	18.6	 	&	0.13		&  &	\ri{9.2}	&	\ri{0.57}		\smallskip		 \\

$20_{*,19}-19_{*,18}$	& 232 736.681 &		9	&	3	&	3.37		 &  &		2.7		&	2.1		&	1.12			& &	19.8	 	&	0.13		&  &	\ri{8.8}	&	\ri{0.54}					 \\
$21_{*,21}-20_{*,20}$ 	& 232 836.881 &		9	&	3	&	3.70		 &  &		2.7		&	2.1		&	1.23			& &	19.8	 	&	0.14		&  &	\ri{8.8}	&	\ri{0.59}		\smallskip		 \\
	
$21_{*,20}-20_{*,19}$	& 243 565.804 &		\bi{2}	&	\bi{2}	&	\bi{0.88}	 &  &		2.7		&	2.1		&	1.18			& &	16.9	 	&	0.13		&  &	\ri{8.5}	&	\ri{0.56}					 \\
$22_{*,22}-21_{*,21}$	& 243 666.100 &		\bi{2}	&	\bi{2}	&	\bi{0.96}	 &  &		2.7		&	2.1		&	1.30			& &	16.9	 	&	0.14		&  &	\ri{8.5}	&	\ri{0.60}		\smallskip		 \\
	
$22_{*,21}-21_{*,20}$	& 254 394.200 &		\bi{2}	&	\bi{2}	&	\bi{0.90}	 &  &		2.7		&	2.1		&	1.22			& &	16.2	 	&	0.14		&  &	\ri{8.1}	&	\ri{0.57}					 \\
$23_{*,23}-22_{*,22}$	& 254 494.539 &		\bi{2}	&	\bi{2}	&	\bi{1.00}	 &  &		2.7		&	2.1		&	1.32			& &	16.2	 	&	0.15		&  &	\ri{8.1}	&	\ri{0.62}		\smallskip		 \\
\enddata
\tablecomments{A zeroth-order best fit was achieved for all telescopes and lines of  $N_T$ = $6 \times 10^{14}$ cm$^{-2}$ and $T_{rot}$ = 80 K for column density and rotational temperature with the exception of the 102 GHz transitions which are best fit with  $N_T$ = $4 \times 10^{14}$ cm$^{-2}$ and $T_{rot}$ = 80 K.  Predictions shown in \ri{red italics} are for transitions occurring within the observable frequencies covered by the facility indicated, but which were not observed in this work.  These predictions were generated using the methods outlined in \S\ref{dataanalysis} and using the parameters given in this table.}
\tablenotetext{a}{Assumes a 2$\arcsec$ source size.}
\tablenotetext{b}{Assumes a 3$\arcsec$ source size.}
\tablenotetext{c}{Assumes a 3.25$\arcsec$ source size.}
\tablenotetext{d}{Predictions in \bi{blue italics} indicate a prediction made for the ALMA facility rather than BIMA which is no longer in service.}
\tablenotetext{e}{Predicted beam sizes are for synthesized beam for CARMA 15 C-array configuration}
\tablenotetext{f}{Predictions for 221, 232, 243, and 254 GHz lines are not corrected for main beam efficiency.}
\label{models}
\end{deluxetable}

\appendix
\section*{Appendix A}
\label{appendixa}

In our case, urea is assumed, under LTE conditions, to be a compact core with no significant velocity gradient and with a system velocity close to that of Sgr B2 (N-LMH) (64 km s$^{-1}$ ). In addition, all array observations were centered at the same region where we could readily derive the correlation coefficients with flux densities at the same velocity plane. By comparing maps of the same velocity channel, the uncertainty introduced by kinematic differences is mostly reduced. The correlation coefficient $r$ between a given pair of urea lines, is defined in Equation \ref{rvalue}.
\begin{equation}
r=\frac{\Sigma(x_i-\bar{x})(y_i-\bar{y})}{\sqrt{\Sigma (x_i-\bar{x})^2}\sqrt{\Sigma (y_i-\bar{y})^2}}
\label{rvalue}
\end{equation}
where $x_i$ and $y_i$ are flux densities of the individual pixel in the same small box bounding the central source in the line maps.  We have calculated the significance of the correlation using the Student's $t$ test.  The $t$ test is only valid when the variables are normally distributed and the variation in each variable is similar.  Since we can fit the sources with Gaussian beam, and the noise levels are equivalent in all maps, the $t$ test is applicable to the flux densities.  The statistical test value $t$ based on $r$ is given by Equation \ref{tvalue} with $n-2$ degrees of freedom \citep{Neter1996}.
\begin{equation}
t=\frac{r\sqrt{n-2}}{\sqrt{1-r^2}}
\label{tvalue}
\end{equation}
Since the smallest component resolved in a map is a beam (hence the units of Jy per beam for flux densities), the number of points $n$ should be the number of beams that filled the area of interest instead of the total number of pixels.  The box chosen for all the lines ($7\arcsec \times 6\arcsec$) yielded $(10-2)=8$ degrees of freedom. Finally, the cumulative density function of the Student's $t$ distribution is used to determine the confidence interval of the resulting test statistics.

Table \ref{ttest} lists the correlation coefficient $r$, test value $t$, and the confidence interval between any two transitions. Compared to the critical $r(95\%)=0.632$, the results suggest some correlation between these lines. Lines contaminated by interlopers but still peaked at the same place tend to show more correlation with the others. For example, 222.007 GHz shows an average $r=0.677$, suggesting good correlation with all other lines. This is not a surprise considering it has a bright large contour in the map. On the other hand, the 232.837 GHz line is blended with some unidentified line at 64 km s$^{-1}$ with a slightly different distribution, as can be seen clearly in the channel maps; its correlation with all the other lines are below 90\%, except with 222.007 GHz and 254.495 GHz, both of which are more extended due to contamination by nearby transitions. Incidentally, 243.566 and 243.666 GHz seem to peak slightly north than the other six lines, resulting in a lower $r$ in most cases. The shift of the peaks is also seen in the continuum emission in the same track, suggesting a possible artificial effect caused by phase calibration. In general, we found that $r$ is more sensitive to the peak location than to the size of the emission contour appearing on the map. This behavior is supported by other known transitions we tested, e.g. CH$_3$OH at 247.968 GHz and CH$_3$CCH at 222.014 GHz. Both lines are bright and have contours as large as the box we chose, and they both correlate very well $(>$99.5\%) with the 232.737 GHz and with each other. Since both CH$_3$OH and CH$_3$CCH originate from the LMH hot core and are not as compact as urea, it is understandable that they are well-correlated. In addition, a modeled Gaussian emission generated with the same attributes as the 232.737 GHz line Þt but a slight offset to the southeast of the peak resulted in very little correlation $(r=0.256, 47.5\%)$. Overall, all of the observed urea lines correlate with each other  to the extent that they peak around the same region, indicating a common origin. 

The correlation model does not decide exactly where the emission originates, but instead, it gives information on how well the emission sources associate in space. Although there is some problem when multiple comparisons are considered in statistics, namely, at least one statistical significant conclusion at the 0.05 level will occur by chance in every 20 comparisons, a correction is not necessary in our case for the following reasons: First, the image fidelity always remains an uncertainty for array observation. Due to the limitation of the CARMA instrument, the 243 and 254 GHz observations are at the high end of the correlator bands and therefore the frequency locks sometimes fail on a certain antenna, causing phase jump during observation. Weather is another crucial factor for such high frequency observation. A slight change in opacity in a short period of time may result in unreliable gain solution. These influences usually do not greatly affect the continuum but could affect the peak location in the channel maps. Even if we had the best data calibration, we could not quantify and remove this uncertainty from the maps. Second, the LMH hot core is an active star-forming region with outflows and ultra compact H II regions impacting one another. Only a handful of molecules have been studied for kinematic distributions. The statistical significance of the correlation, even from the same molecule, might not achieve the conventional value of 0.05 after all. A corrected significance level for multiple-comparisons hypothesis testing could be too conservative and rule out most correlated pairs that could be perceived by eye. Since our intention of applying statistics is to provide a quantitative aspect of spatial correlation, such stringent hypothesis testing may not be as constructive for data interpretation.

\begin{deluxetable}{cccc}

\tabletypesize{\scriptsize}
\tablecaption{Statistical correlation between flux densities of urea transitions.\label{tab:ttest}}
\tablenum{A1}
\tablewidth{0pt}
\tablehead{
         \colhead{} & 
         \colhead{Correlation} &
         \colhead{\tablenotemark{a}Student's $t$} &
         \colhead{Confidence} \\
         \colhead{Transitions} & 
         \colhead{Coefficient $r$} &
         \colhead{Test Value $t$} &
         \colhead{Interval (\%)} }
\startdata
221.907/222.007 & 0.713 & 2.875 & 97.9 \\
\phm{221.907}/232.737 & 0.724 & 2.969 & 98.2 \\
\phm{221.907}/232.837 & 0.423 & 1.319 & 77.6 \\
\phm{221.907}/243.566 & 0.424 & 1.323 & 77.8 \\
\phm{221.907}/243.666 & 0.311 & 0.925 & 61.8 \\
\phm{221.907}/254.394 & 0.452 & 1.435 & 81.1 \\
\phm{221.907}/254.495 & 0.389 & 1.194 & 73.3 \\
222.007/232.737 & 0.884 & 5.335 & 99.9  \\
\phm{221.907}/232.837 & 0.582 & 2.026 & 92.3 \\
\phm{221.907}/243.566 & 0.730 & 3.020 & 98.3 \\
\phm{221.907}/243.666 & 0.512 & 1.686 & 86.9 \\
\phm{221.907}/254.394 & 0.550 & 1.864 & 90.1 \\
\phm{221.907}/254.495 & 0.769 & 3.403 & 99.1 \\
232.737/232.837 & 0.512 & 1.687 & 87.0 \\
\phm{221.907}/243.566 & 0.642 & 2.370 & 95.5 \\
\phm{221.907}/243.666 & 0.535 & 1.792 & 88.9 \\
\phm{221.907}/254.394 & 0.465 & 1.485 & 82.4 \\
\phm{221.907}/254.495 & 0.766 & 3.370 & 99.0 \\
232.837/243.566 & 0.496 & 1.615 & 85.5 \\
\phm{221.907}/243.666 & 0.525 & 1.743 & 88.1 \\
\phm{221.907}/254.394 & 0.469 & 1.500 & 82.8 \\
\phm{221.907}/254.495 & 0.581 & 2.017 & 92.2 \\
243.566/243.666 & 0.436 & 1.369 & 79.2 \\
\phm{221.907}/254.394 & 0.309 & 0.920 & 61.5 \\
\phm{221.907}/254.495 & 0.439 & 1.381 & 79.5 \\
243.666/254.394 & 0.252 & 0.735 & 51.7 \\
\phm{221.907}/254.495 & 0.479 & 1.545 & 83.9 \\
254.394/254.495 & 0.484 & 1.564 & 84.4 \\
\enddata
\tablenotetext{a}{Degree of freedom $df$(n-2)=8.}
\label{ttest}
\end{deluxetable}

\clearpage

\section*{Appendix B}
\label{appendixb}

\begin{deluxetable}{l r}
\tablecaption{Rotational analysis of Urea}
\tablenum{B1}
\tablewidth{0pt}
\tablehead{
	\colhead{Parameter} &
	\colhead{Value}
}
\startdata
$A$ /MHz			&	11 233.3213(10)	\\
$B$ /MHz			&	10 369.3727(11)	\\
$C$ /MHz			&	5 416.6320(9)		\\
$\Delta_J$ /kHz		&	5.5268(21)		\\
$\Delta_{JK}$ /kHz	&	-5.2785(85)		\\
$\Delta_K$ /kHz		&	10.961(11)		\\
$\delta_J$ /kHz		&	2.40058(79)		\\
$\delta_K$ /kHz		&	3.9044(37)		\\
$\sigma$ (weighted)	&	1.173			\\
\enddata
\label{franksparams}
\tablecomments{Uncertainties are given in parentheses in units of the last significant digit and are of Type A with coverage factor k =2 (Taylor \& Kuyatt 1994).}
\end{deluxetable}

\begin{deluxetable}{r r r r r r c c c}
\tabletypesize{\scriptsize}
\tablecaption{Measurements and fit of the hyperfine-free urea transitions}
\tablenum{B2}
\tablewidth{0pt}
\tablehead{
         \colhead{$J^{\prime}$} & 
         \colhead{$K_a^{\prime}$} &
         \colhead{$K_c^{\prime}$} &
         \colhead{$J^{\prime\prime}$} &
         \colhead{$K_a^{\prime\prime}$} & 
         \colhead{$K_c^{\prime\prime}$} &
         \colhead{Frequency (MHz)} &
         \colhead{Obs. - Calc.} &
         \colhead{Ref.} 
}
\startdata
  1&   1&   0 &   1&   0&   1&      5816.667( 10)&        -0.0044&  Kas86\\
  2&   2&   0 &   2&   1&   1&      6784.070( 40)&         0.0389&  Bro75\\
  3&   3&   0 &   3&   2&   1&      8415.222(  2)&         0.0011&  Kre96\\
  5&   4&   1 &   5&   3&   2&     13395.250( 40)&         0.0049&  Bro75\\
  4&   3&   1 &   4&   2&   2&     13517.386(  2)&         0.0018&  Kre96\\
  6&   5&   1 &   6&   4&   2&     14015.320( 40)&         0.0111&  Bro75\\
  3&   2&   1 &   3&   1&   2&     14138.687(  2)&         0.0009&  Kre96\\
  5&   5&   0 &   5&   4&   1&     14222.390( 40)&         0.0468&  Bro75\\
  2&   1&   1 &   2&   0&   2&     14961.487(  2)&         0.0012&  Kre96\\
  7&   6&   1 &   7&   5&   2&     15556.720( 40)&         0.0037&  Bro75\\
  1&   1&   1 &   0&   0&   0&     16649.945(  2)&        -0.0032&  Kre96\\
  2&   2&   1 &   2&   1&   2&     17449.887(  1)&        -0.0015&  Kre96\\
  8&   7&   1 &   8&   6&   2&     18136.359( 40)&         0.0101&  Bro75\\
  6&   6&   0 &   6&   5&   1&     18411.801( 40)&         0.0243&  Bro75\\
  3&   3&   1 &   3&   2&   2&     18806.682(  2)&         0.0023&  Kre96\\
 10&   8&   2 &  10&   7&   3&     20697.650( 40)&         0.0133&  Bro75\\
  9&   7&   2 &   9&   6&   3&     20287.699( 40)&         0.0993&  Bro75\\
  8&   6&   2 &   8&   5&   3&     20800.600( 40)&         0.0635&  Bro75\\
  9&   8&   1 &   9&   7&   2&     21771.500( 40)&        -0.0087&  Bro75\\
 10&   9&   1 &  10&   8&   2&     26332.801( 40)&        -0.0212&  Bro75\\
 13&  10&   3 &  13&   9&   4&     26813.029( 40)&        -0.0464&  Bro75\\
 14&  11&   3 &  14&  10&   4&     27071.240( 40)&        -0.0384&  Bro75\\
  2&   1&   2 &   1&   0&   1&     27483.199( 40)&         0.0279&  Bro75\\
 12&   9&   3 &  12&   8&   4&     27638.520( 40)&         0.0358&  Bro75\\
  5&   4&   2 &   5&   3&   3&     27760.000( 40)&         0.0413&  Bro75\\
  8&   8&   0 &   8&   7&   1&     28346.551( 40)&        -0.0022&  Bro75\\
 15&  12&   3 &  15&  11&   4&     28611.721( 40)&         0.0009&  Bro75\\
 13&  11&   2 &  13&  10&   3&     28881.199( 40)&         0.0260&  Bro75\\
  7&   4&   3 &   7&   3&   4&     36079.148( 40)&        -0.0021&  Bro75\\
  6&   3&   3 &   6&   2&   4&     36840.801( 40)&         0.0734&  Bro75\\
  5&   2&   3 &   5&   1&   4&     37264.230( 40)&         0.0877&  Bro75\\
  6&   4&   3 &   6&   3&   4&     37441.148( 40)&        -0.0310&  Bro75\\
  7&   5&   3 &   7&   4&   4&     37482.699( 40)&         0.0765&  Bro75\\
  8&   6&   3 &   8&   5&   4&     37671.199( 40)&         0.1258&  Bro75\\
  3&   0&   3 &   2&   1&   2&     37815.398( 40)&         0.0747&  Bro75\\
 20&  16&   4 &  20&  15&   5&     37896.852( 40)&        -0.0153&  Bro75\\
  3&   1&   3 &   2&   0&   2&     37926.699( 40)&         0.0758&  Bro75\\
  9&   7&   3 &   9&   6&   4&     38092.340( 40)&         0.0113&  Bro75\\
 10&  10&   0 &  10&   9&   1&     38627.719( 40)&         0.0354&  Bro75\\
 14&  10&   4 &  14&   9&   5&     38635.352( 40)&         0.0282&  Bro75\\
 10&   8&   3 &  10&   7&   4&     38833.500( 40)&        -0.0023&  Bro75\\
  2&   2&   1 &   1&   1&   0&     39116.398( 40)&         0.0099&  Bro75\\
 21&  16&   5 &  21&  15&   6&     39175.680( 40)&         0.0319&  Bro75\\
 22&  17&   5 &  22&  16&   6&     39238.930( 40)&         0.0153&  Bro75\\
  2&   2&   0 &   1&   1&   1&     48261.301( 40)&         0.0620&  Bro75\\
  4&   0&   4 &   3&   1&   3&     48697.578( 40)&         0.1556&  Bro75\\
  4&   1&   4 &   3&   0&   3&     48705.852( 40)&         0.1239&  Bro75\\
 17&  12&   5 &  17&  11&   6&     48733.680( 40)&        -0.0016&  Bro75\\
  4&   1&   3 &   3&   2&   2&     59185.316( 50)&        -0.0186&  NIST \\
  4&   2&   3 &   3&   1&   2&     59747.449( 20)&        -0.0859&  NIST \\
  3&   3&   1 &   2&   2&   0&     61972.109( 30)&         0.0303&  NIST \\
  4&   2&   2 &   3&   3&   1&     66219.094( 60)&        -0.0195&  NIST \\
  3&   3&   0 &   2&   2&   1&     69403.570( 40)&         0.0495&  NIST \\
  5&   1&   4 &   4&   2&   3&     70251.141( 20)&        -0.0289&  NIST \\
  5&   2&   4 &   4&   1&   3&     70309.844( 40)&         0.0248&  NIST \\
  6&   0&   6 &   5&   1&   5&     70366.531( 30)&         0.0259&  NIST \\
  6&   1&   6 &   5&   0&   5&     70366.531( 30)&        -0.0033&  NIST \\
  4&   3&   2 &   3&   2&   1&     72858.781( 60)&        -0.0264&  NIST \\
 17&  10&   7 &  17&   9&   8&     78340.031( 50)&        -0.0542&   IRA \\
 17&  11&   7 &  17&  10&   8&     78460.172( 50)&        -0.0265&   IRA \\
 16&   9&   7 &  16&   8&   8&     78844.586( 50)&        -0.0187&   IRA \\
 16&  10&   7 &  16&   9&   8&     78896.742( 50)&        -0.0374&   IRA \\
 26&  22&   5 &  26&  21&   6&     79165.633( 50)&         0.0429&   IRA \\
 15&   8&   7 &  15&   7&   8&     79244.750( 50)&        -0.0871&   IRA \\
 15&   9&   7 &  15&   8&   8&     79265.953( 50)&        -0.0205&   IRA \\
 14&   7&   7 &  14&   6&   8&     79563.992( 50)&        -0.0137&   IRA \\
 14&   8&   7 &  14&   7&   8&     79571.945( 50)&         0.0289&   IRA \\
 13&   6&   7 &  13&   5&   8&     79818.312( 50)&        -0.0215&   IRA \\
 13&   7&   7 &  13&   6&   8&     79821.008( 50)&        -0.0258&   IRA \\
  5&   2&   3 &   4&   3&   2&     80265.312( 40)&        -0.0416&  NIST \\
 26&  18&   8 &  26&  17&   9&     80455.242( 50)&         0.0498&   IRA \\
  6&   1&   5 &   5&   2&   4&     81104.133( 20)&         0.0020&  NIST \\
  6&   2&   5 &   5&   1&   4&     81108.773( 40)&        -0.0408&  NIST \\
  7&   0&   7 &   6&   1&   6&     81199.203( 20)&         0.0037&  NIST \\
  7&   1&   7 &   6&   0&   6&     81199.203( 20)&         0.0021&  NIST \\
  5&   3&   3 &   4&   2&   2&     81942.656( 20)&        -0.0238&  NIST \\
  4&   4&   1 &   3&   3&   0&     85082.469( 30)&         0.0408&  NIST \\
  6&   2&   4 &   5&   3&   3&     91769.187( 80)&        -0.0909&  NIST \\
  6&   3&   4 &   5&   2&   3&     92004.570( 60)&         0.0423&  NIST \\
  8&   0&   8 &   7&   1&   7&     92031.820( 40)&         0.0072&  NIST \\
  8&   1&   8 &   7&   0&   7&     92031.820( 40)&         0.0071&  NIST \\
  5&   4&   2 &   4&   3&   1&     96185.227( 30)&        -0.0272&  NIST \\
  4&   3&   1 &   3&   2&   2&     98543.320( 90)&         0.1426&  NIST \\
  6&   3&   3 &   5&   4&   2&    100850.008( 30)&        -0.0389&  NIST \\
  7&   2&   5 &   6&   3&   4&    102680.070( 80)&         0.0005&  NIST \\
  7&   3&   5 &   6&   2&   4&    102703.578( 10)&         0.0075&  NIST \\
  8&   1&   7 &   7&   2&   6&    102767.562( 40)&         0.0375&  NIST \\
  8&   2&   7 &   7&   1&   6&    102767.562( 40)&         0.0173&  NIST \\
  9&   0&   9 &   8&   1&   8&    102864.320( 60)&         0.0108&  NIST \\
  9&   1&   9 &   8&   0&   8&    102864.320( 60)&         0.0108&  NIST \\
  6&   4&   3 &   5&   3&   2&    104665.937( 30)&         0.0147&  NIST \\
  5&   5&   1 &   4&   4&   0&    108265.133( 20)&        -0.0057&  NIST \\
  5&   5&   0 &   4&   4&   1&    112023.898( 20)&         0.0133&  NIST \\
  9&   1&   8 &   8&   2&   7&    113599.102( 60)&         0.0364&  NIST \\
  9&   2&   8 &   8&   1&   7&    113599.102( 60)&         0.0352&  NIST \\
 10&   0&  10 &   9&   1&   9&    113696.602( 60)&        -0.0558&  NIST \\
 10&   1&  10 &   9&   0&   9&    113696.602( 60)&        -0.0558&  NIST \\
  7&   4&   4 &   6&   3&   3&    113895.820( 40)&        -0.0397&  NIST \\
 10&   1&   9 &   9&   2&   8&    124430.687( 50)&        -0.0118&   IRA \\
 11&   0&  11 &  10&   1&  10&    124528.852( 50)&         0.0177&   IRA \\
 26&  14&  12 &  26&  13&  13&    131766.641( 70)&         0.0828&   IRA \\
 26&  15&  12 &  26&  14&  13&    131766.641( 70)&        -0.0682&   IRA \\
 24&  12&  12 &  24&  11&  13&    132439.187( 70)&        -0.0478&   IRA \\
 24&  13&  12 &  24&  12&  13&    132439.187( 70)&        -0.0682&   IRA \\
 23&  11&  12 &  23&  10&  13&    132718.578( 50)&        -0.0665&   IRA \\
 22&  11&  12 &  22&  10&  13&    132964.703( 50)&        -0.0749&   IRA \\
 21&   9&  12 &  21&   8&  13&    133180.641( 50)&        -0.1150&   IRA \\
 20&   9&  12 &  20&   8&  13&    133369.406( 50)&        -0.0538&   IRA \\
 19&   7&  12 &  19&   6&  13&    133533.453( 50)&        -0.0840&   IRA \\
 18&   6&  12 &  18&   5&  13&    133675.516( 50)&         0.0775&   IRA \\
 17&   6&  12 &  17&   5&  13&    133797.469( 50)&         0.0348&   IRA \\
 16&   4&  12 &  16&   3&  13&    133901.641( 50)&         0.0113&   IRA \\
 15&   4&  12 &  15&   3&  13&    133990.000( 50)&         0.0214&   IRA \\
 14&   3&  12 &  14&   2&  13&    134064.344( 50)&         0.0513&   IRA \\
  9&   3&   6 &   8&   4&   5&    135099.266( 50)&        -0.0067&   IRA \\
  9&   4&   6 &   8&   3&   5&    135107.672( 50)&        -0.0142&   IRA \\
 10&   2&   8 &   9&   3&   7&    135168.500( 50)&         0.0207&   IRA \\
 12&   1&  12 &  11&   0&  11&    135360.812( 50)&        -0.0056&   IRA \\
 27&  14&  13 &  27&  13&  14&    142671.297( 70)&         0.1508&   IRA \\
 27&  15&  13 &  27&  14&  14&    142671.297( 70)&         0.1338&   IRA \\
 28&  15&  13 &  28&  14&  14&    142314.953( 70)&         0.1739&   IRA \\
 28&  16&  13 &  28&  15&  14&    142314.953( 70)&         0.1286&   IRA \\
 26&  14&  13 &  26&  13&  14&    142990.187( 50)&         0.0022&   IRA \\
 25&  12&  13 &  25&  11&  14&    143275.047( 50)&        -0.0296&   IRA \\
 24&  12&  13 &  24&  11&  14&    143528.719( 50)&        -0.0562&   IRA \\
 23&  10&  13 &  23&   9&  14&    143753.922( 50)&        -0.0547&   IRA \\
 22&  10&  13 &  22&   9&  14&    143953.219( 50)&         0.0359&   IRA \\
 21&   8&  13 &  21&   7&  14&    144128.672( 50)&        -0.0404&   IRA \\
 20&   8&  13 &  20&   7&  14&    144282.609( 50)&        -0.1112&   IRA \\
 19&   6&  13 &  19&   5&  14&    144417.172( 50)&        -0.0417&   IRA \\
 18&   6&  13 &  18&   5&  14&    144534.016( 50)&        -0.0486&   IRA \\
 17&   4&  13 &  17&   3&  14&    144634.984( 50)&        -0.0352&   IRA \\
 16&   4&  13 &  16&   3&  14&    144721.734( 50)&         0.0272&   IRA \\
 15&   3&  13 &  15&   2&  14&    144795.703( 50)&         0.0560&   IRA \\
 14&   1&  13 &  14&   0&  14&    144858.359( 70)&         0.1038&   IRA \\
 14&   2&  13 &  14&   1&  14&    144858.359( 70)&         0.1038&   IRA \\
 10&   3&   7 &   9&   4&   6&    145919.906( 50)&         0.1399&   IRA \\
 11&   2&   9 &  10&   3&   8&    145998.094( 50)&        -0.0106&   IRA \\
 12&   2&  11 &  11&   1&  10&    146093.797( 50)&         0.0406&   IRA \\
 13&   0&  13 &  12&   1&  12&    146192.547( 50)&        -0.0409&   IRA \\
 17&   1&  16 &  16&   2&  15&    200247.812( 50)&        -0.0158&   IRA \\
 18&   0&  18 &  17&   1&  17&    200347.609( 50)&         0.0073&   IRA \\
 17&   2&  15 &  16&   3&  14&    210977.687( 50)&         0.0470&   IRA \\
 18&   1&  17 &  17&   2&  16&    211077.766( 50)&        -0.0352&   IRA \\
 19&   0&  19 &  18&   1&  18&    211177.672( 50)&        -0.0398&   IRA \\
 14&   6&   8 &  13&   7&   7&    221532.422( 50)&        -0.0697&   IRA \\
 14&   7&   8 &  13&   6&   7&    221535.500( 50)&        -0.0312&   IRA \\
 13&   7&   6 &  12&   8&   5&    221540.453( 50)&        -0.0270&   IRA \\
 15&   5&  10 &  14&   6&   9&    221543.047( 50)&         0.0621&   IRA \\
 16&   4&  12 &  15&   5&  11&    221615.766( 50)&         0.0298&   IRA \\
 17&   3&  14 &  16&   4&  13&    221708.078( 50)&         0.0363&   IRA \\
 18&   2&  16 &  17&   3&  15&    221806.922( 50)&         0.0271&   IRA \\
 19&   1&  18 &  18&   2&  17&    221907.484( 50)&         0.0398&   IRA \\
 20&   0&  20 &  19&   1&  19&    222007.422( 50)&        -0.0611&   IRA \\
 15&   6&   9 &  14&   7&   8&    232324.344(500)&         0.1315&   IRA \\
 15&   7&   9 &  14&   6&   8&    232324.344(500)&        -0.1577&   IRA \\
 16&   5&  11 &  15&   6&  10&    232361.344( 50)&         0.0111&   IRA \\
 17&   4&  13 &  16&   5&  12&    232441.437( 50)&         0.0059&   IRA \\
 18&   3&  15 &  17&   4&  14&    232536.109( 50)&         0.0041&   IRA \\
 20&   1&  19 &  19&   2&  18&    232736.719( 50)&        -0.0184&   IRA \\
 21&   0&  21 &  20&   1&  20&    232836.906( 50)&         0.0079&   IRA \\
\enddata
\tablerefs{\\Bro75: Brown, R. D., Godfrey, P. D., \& Storey, J. 1975, J. Mol. Spectrosc. 58, 445.\\
Kas86: Kasten, W. \& Dreizler, H. 1986, Z. Naturforsch. 41a, 1173.\\
Kre96: Kretschmer, U., Consalvo, D., Knaack, A. Shade, W. Stahl, W. \& Dreizler, H. 1996, Mol. Phys. 87, 1159.\\
NIST:  Present work, measured at the National Institute of Standards and Technology\\
IRA:  Present work, measured at the Institute of Radio Astronomy of the NASU}
\label{franks}
\end{deluxetable}

\end{document}